\documentclass[aps,twocolumn,jcp]{revtex4-1}
\usepackage{graphics,graphicx}
\usepackage{amsmath}
\begin{document}
\newcommand{\bstfile}{aps} %alternative styles: osa, prasty or revtex
\newcommand{\bibs}{d:/Dad/Bibliography/TexBiB/final}

\title{The effect of electron interactions on the universal properties of systems with optimized off-resonant intrinsic hyperpolarizability}
\author{David S. Watkins* and Mark G. Kuzyk}\email{kuz@wsu.edu}
\address{Department of Physics and Astronomy, Washington State University, Pullman, Washington  99164-2814; and *Department of Mathematics}

\begin{abstract}
Because of the potentially large number of important applications of nonlinear optics, researchers have expended a great deal of effort to optimize the second-order molecular nonlinear-optical response, called the hyperpolarizability.  The focus of our present studies is the {\em intrinsic} hyperpolarizability, which is a scale-invariant quantity that removes the effects of simple scaling, thus being the relevant quantity for comparing molecules of varying sizes.  Past theoretical studies have focused on structural properties that optimize the intrinsic hyperpolarizability, which have characterized the structure of the quantum system based on the potential energy function, placement of nuclei, geometry, and the effects of external electric and magnetic fields.  Those previous studies focused on single-electron models under the influence of an average potential.  In the present studies, we generalize our calculations to two-electron systems and include electron interactions. As with the single-electron studies, universal properties are found that are common to all systems -- be they molecules, nanoparticles, or quantum gases -- when the hyperpolarizability is near the fundamental limit.
\end{abstract}

\maketitle

\section{Introduction}

Materials with large nonlinear-optical susceptibilities are central for optical applications such as telecommunications,\cite{wang04.01} three-dimensional nano-photolithography,\cite{cumps99.01,kawat01.01} and making new materials\cite{karot04.01} for novel cancer therapies.\cite{roy03.01} Semi-empirical theoretical modeling of complex molecules, with feedback from experiments, has resulted in a better understanding of how to make materials with larger nonlinear-optical response.\cite{Breda94.01,Albot98.01}

Accurate theoretical modeling of a variety of specific molecules is a critical part in the development of structure-property relationships.\cite{fu07.01,Rumi00.01,liao05.01,nandi08.01} In contrast, our work seeks to build a {\em broad} theoretical understanding of the nonlinear-optical response that applies to all quantum systems,\cite{kuzyk09.01} of which molecules are a small subset.  The uniqueness of this approach is that it seeks to identify universal properties that are common to all systems sharing a nonlinear response that is close to the fundamental limit.\cite{kuzyk10.01} In the present work, we show that interactions between electrons do not have an effect on the universal properties, and that scaling due to the addition of an electron is as predicted from quantum limit theory.

For electric fields that are small compared with internal molecular fields, the induced dipole moment of a molecule, $p$, can be expressed as a series in the applied electric field, $E$,
\begin{equation}\label{dipole-expand}
p = \alpha E + \beta E^2 + \gamma E^3 + \dots,
\end{equation}
where $\alpha$ is the polarizability, $\beta$, the hyperpolarizability, $\gamma$ the second hyperpolarizability, and so on.  In general, these coefficients are tensors that depend on the frequencies of the applied electric field.  However, our studies focus on (1) the non-resonant regime, were all frequencies are small compared with the Bohr frequencies of the molecule; and, (2) the $xxx$ tensor component of $\beta$, where $x$ is the axis along which the hyperpolarizability is the largest.

We consider only the {\em electronic hyperpolarizability in the off-resonant limit} - where the nuclei are assumed not to contribute.  Chernyak and coworkers showed that in second and third harmonic generation experiments, when the wavelengths are tuned below the excited electronic state energy and above the nuclear excitation energies, the purely electronic hyperpolarizabilities account for 90-95\% of the total.\cite{chern20.01} Our theoretical analysis and experimental results come to the same conclusion.\cite{Tripa04.01,tripa07.01} Bishop and coworkers point out that while this may be true for harmonic generation hyperpolarizabilties, vibronic contributions can be dominant when any of the frequencies vanish or in self action effects where negative frequencies cancel positive ones.\cite{bisho20.01} The literature is extensive on the topic of conditions when the vibronic contributions are dominant.\cite{das93.01,kirtm97.01,paine98.01,bisho98.01}

In the theoretical studies presented here, we focus specifically on the zero-frequency limit of the electronic response.  Our results can be compared with second-harmonic generation measurements of the off-resonant hyperpolarizability, which is dominated by the electronic response, upon extrapolation to zero frequency\cite{clays91.01} using a two-level dispersion model.\cite{oudar77.03}  All future references to hyperpolarizabilities refer to the first {\em electronic} hyperpolarizability at zero-frequency.  We stress that our goal is to understand the zero-frequency electronic response, independent of the subtleties associated with measuring it.

The hyperpolarizability can be calculated for any quantum system with the dipole energy due to the applied electric field as a perturbation.  This results in a sum-over-states (SOS) expansion,\cite{orr71.01} which is expressed in terms of the position matrix elements and energy eigenvalues of the unperturbed molecule.

The fundamental limit of the hyperpolarizability is defined as the upper bound determined from fundamental quantum principles, without approximation.  Thus, the fundamental limit of the hyperpolarizability is an absolute quantity that applies to any quantum system that is described by the Schr\"{o}dinger Equation.  Rather than being a single number, the limit is a function of the energy difference between ground and first excited state, $E_{10} = E_1 - E_0$, and the number of electrons.

This upper bound is calculated using the Thomas-Reiche-Kuhn sum rules,\cite{Bethe77.01} which are exact and follow directly from the Schr\"{o}dinger equation.  They relate the energy eigenvalues and position matrix elements to each other, and are used to simplify the SOS expansion of the hyperpolarizability.\cite{orr71.01} This approach was applied to the calculation the fundamental limit of the off-resonance hyperpolarizability, which yields\cite{kuzyk00.01}
\begin{equation}\label{limit}
\beta_{MAX} = \sqrt[4]{3} \left( \frac {e \hbar} {\sqrt{m}}
\right)^3 \cdot \frac {N^{3/2}} {E_{10}^{7/2}} ,
\end{equation}
where $N$ is the number of electrons, $E_{10}$ the energy difference between the
first excited state and the ground state, $E_{10} = E_1 - E_0$, and $m$ is the electron mass.  The fundamental limit of the second hyperpolarizability has also been calculated.\cite{kuzyk00.02} These calculations show that there is a large gap between the hyperpolarizabilities and the fundamental limit.\cite{kuzyk00.02,kuzyk01.01,kuzyk00.01,kuzyk03.01,kuzyk03.02,kuzyk04.02}  The existence of such a limit is both scientifically interesting and of practical utility because it provides a target for making optimized materials.

One approach to the development of structure-property relationships is to correlate the nonlinear-optical response with variations in the structure.  However, it is difficult to decouple structural properties from simple scaling - i.e. the change in the effective size of a molecule with a change in its structure.  It can be shown that the intrinsic hyperpolarizability $\beta_{int}$, which is obtained by dividing the actual hyperpolarizability $\beta$ by the fundamental limit $\beta_{MAX}$:
\begin{equation}\label{intrinsic-beta}
\beta_{int} = \beta / \beta_{MAX} ,
\end{equation}
is invariant to simple scaling.\cite{kuzyk10.01}  A comparison of the intrinsic hyperpolarizabilties of molecules can be used to determine the properties that are relevant for optimizing the nonlinear-optical response.

Past studies have considered the optimization of the intrinsic hyperpolarizability by varying the shape of the potential energy function,\cite{zhou06.01,zhou07.02} which lead to the new paradigm of large-$\beta$ molecules using modulation of conjugation for which there is some experimental evidence;\cite{perez07.01,perez09.01} studying the effect of geometry by varying the charges and positions of point nuclei;\cite{kuzyk06.01} and the effect of applying an arbitrary electromagnetic field.\cite{watkins09.01}  All of these studies find that when the hyperpolarizability of a quantum system is near the fundamental limit, certain universal properties result.  For example, there are many one-dimensional potentials - each of distinct shape, but different from each other --  whose optimized intrinsic hyperpolarizability is 0.709.  For these systems, the ratio of second to first excited state energy is $E \equiv E_{10}/E_{20} \approx 0.48$.  Furthermore, typically over 90\% of the optimized hyperpolarizability is due to two excited states.  Thus, {\em all optimized materials} can be well approximated by a three-level model.

When maximizing the off-resonance hyperpolarizability of a one-dimensional molecule, planar molecules made of point nuclei, or systems with externally applied electric and magnetic fields, one-electron models were unable to get intrinsic hyperpolarizabilities that exceed about 0.708.  However, Monte Carlo simulations that randomly assign energies and position matrix elements under the constraint that the sum rules are obeyed result in $\beta_{int}$ that approaches unity.\cite{kuzyk08.01}  Since such large values have never been observed under any Hamiltonian-based simulation, this suggests that one may need to invoke exotic Hamiltonians to reach the fundamental limit.

The values of $\beta_{int} \approx 0.708$ that we have obtained in our
computer simulations are much higher than any that have been measured
for real molecules.  For example, it was reported that a new molecule with
asymmetric modulation of conjugation has a measured value of $\beta_{int}
= 0.048$ \cite{perez07.01}, but this is still more than a factor of ten
lower than what our simulations have produced.  It is natural to ask whether our simulations give unrealistically large $\beta_{int}$ values simply because we are using one-electron models.\cite{Champ05.01}

For a system of {\em two non-interacting electrons}, the value of $\beta$ is twice that of a system with one electron.  But $\beta_{MAX}$ scales as $2^{3/2}$ due to the factor $N^{3/2}$ in Equation \ref{limit}.  Thus
the highest value of $\beta_{int}$ that we will be able to get with
the intrinsic hyperpolarizability of the two electron system is about $.708/\sqrt{2} \approx
0.5$.  As the number of non-interacting electrons is increased, the intrinsic hyperpolarizability decreases.  For a 10-electron system, this would reduce the intrinsic hyperpolarizability by a factor of 30, the gap observed between the best molecules and the fundamental limit.

In the present work we consider two-electron models and ask whether it
is possible to produce values of $\beta_{int}$ that are substantially
greater than 0.5 when including electron-electron interactions.
We will show that $\beta_{int}$ of the two-electron system is as large
as about 0.708.  Furthermore, we find that the same universal properties hold as in the one-electron case. Thus, the high $\beta_{int}$ values that we obtained in previous work are not simply an artifact of one-electron models.  More importantly, this suggests that one can use simple one-electron models when painting our understanding of the nonlinear-optical response with the broadest of brushes.

\section{Theory}

We use a two-particle Hamiltonian that includes a one-dimensional potential energy function, $V(x)$, that acts on both electrons, and an interaction term between the electrons.  Note that past studies in two dimensions show the same universal behavior as in one dimension, so we find no particular utility in studying these other classes.  If electron interactions do not change the universal properties in 1D, it is unlikely that they will have any effect in 2D.

The potential function is modeled by a piecewise linear function that has 19 degrees of freedom in the computational region.  With 39 degrees of freedom, the results are nearly the same; but, the computations are more time consuming.  We thus use 19 degrees of freedom because it provides a good comprmise between accuracy and speed.

Starting from a given potential function, we use the Nelder-Mead
simplex algorithm \cite{lagar98.01} to vary the shape of the potential
to maximize $\beta_{int}$.  Since there are 19 degrees of freedom, we
are maximizing the hyperpolarizability over a 19-dimensional space.
We have three ways of computing $\beta$ (SOS,\cite{orr71.01} dipole free,\cite{kuzyk05.02} and non-perturbative\cite{zhou07.01}), all of which require solving the Schr\"{o}dinger
eigenvalue problem for the given potential (and in some cases also for
neighboring potentials).  We solve the eigenvalue problem  numerically
using the finite element method.\cite{zienk05.01}

Once we have solved the eigenvalue problem, we can compute transition
moments and then obtain $\beta$ by
the standard Orr and Ward SOS expression
$\beta_{SOS}$ \cite{orr71.01},  the dipole free expression
$\beta_{DF}$ \cite{kuzyk05.02}, or a non-perturbative finite difference approximation
$\beta_{NP}$.\cite{zhou07.01}  In the optimization code we
use $\beta_{NP}$.  That is, we seek to maximize $\beta_{int} =
\beta_{NP}/\beta_{MAX}$.  Once the optimization is complete, we use a comparison of $\beta_{NP}$ with
$\beta_{SOS}$ and $\beta_{DF}$ as a test of convergence.

The exact computation of $\beta_{SOS}$ and $\beta_{DF}$
requires sums over infinitely many states.
We approximate them by summing over the 30 lowest energy
levels.  This is found to give an accurate value for the hyperpolarizability that is almost identical with the result we get when using up to 100 states.\cite{zhou07.02}

To illustrate the difference between the one- and two-electron cases, we take
a closer look at the Schr\"{o}dinger eigenvalue problems.  For the
one-electron problem in one-dimension, we solve
\begin{equation}\label{eq:schroedinger1d}
-\frac{\hbar^{2}}{2m} \frac{\partial^{2}\Psi}{\partial x^{2}} +
V(x)\Psi = E \Psi,
\end{equation}
where the wave function $\Psi$ is a function of $x$.  For two
electrons without interactions, the Schr\"{o}dinger equation is
\begin{equation}\label{eq:schro_nointeract}
-\frac{\hbar^{2}}{2m} \left(\frac{\partial^{2}\Psi}{\partial x_{1}^{2}} +
\frac{\partial^{2}\Psi}{\partial x_{2}^{2}} \right) +
\left( V(x_{1}) + V(x_{2}) \right) \Psi = E \Psi ,
\end{equation}
where $\Psi$ is a function of $x_{1}$ and $x_{2}$.  The solutions of Equation \ref{eq:schro_nointeract} are tensor products of solutions of Equation \ref{eq:schroedinger1d}, and the energy levels are sums of energy
levels of Equation \ref{eq:schroedinger1d}.  The largest $\beta$ values
that we can obtain from Equation \ref{eq:schro_nointeract} are exactly twice
those that we get from Equation \ref{eq:schroedinger1d}, but the largest
value of $\beta_{int}$ is lower by a factor of exactly $\sqrt{2}$
because of the factor $N^{3/2}$ that appears in Equation \ref{limit} for $\beta_{MAX}$.

To include interactions between the electrons, we consider a model of the form
\begin{eqnarray}\label{eq:schro_yesinteract}
&-& \frac {\hbar^{2}} {2m} \nabla^{2}\Psi
+ \left( V(x_{1}) + V(x_{2}) \right. \nonumber \\
&+& \left. V_{2}(\left|x_{1}-x_{2} \right|)\right) \Psi
= E \Psi.
\end{eqnarray}
Here $V(x_i)$ is the external potential that acts on both electrons equivalently.  It can be due to the nuclei or external influences such as an applied electromagnetic field.     $V_{2}(\left|x_{1}-x_{2}\right|)$ is the interaction potential between electrons, which is a sum of two components
$$V_{2}(\left|x_{1}-x_{2}\right|) = V_{21}(\left|x_{1}-x_{2}\right|) +
V_{22}(\left|x_{1}-x_{2}\right|),$$
where $V_{21}$ and $V_{22}$ denote electrostatic repulsion between the electrons and spin
interaction terms, respectively.  These terms are piecewise linear, which is appropriate for a one-dimensional model.

Defining $d = \left|x_{1}-x_{2}\right|$, we assume that the interaction potentials take the form
\begin{equation}\label{eq:v21}
V_{21}(d) = \hbar^{2} \cdot \max\left\{ C_{1} \frac{L-d}{L}, 0 \right\},
\end{equation}
where $L$ is the length of the molecule (we use $L=25$ for all calculations); and,
\begin{equation}\label{eq:v22}
V_{22}(d) = S \cdot \max\left\{ C_{2} \frac{\hat{L}-d}{\hat{L}}, 0 \right\},
\end{equation}
where $\hat{L}= C_{3}L$  with $C_{3} \ll 1$
to ensure that the spin interaction acts only at close range.  Note that this is a one-dimensional coulomb potential that satisfies the general condition $\nabla^2 V(x_1 - x_2) \propto \delta(x_1 - x_2)$, where $\delta(x_1 - x_2)$ is the Dirac delta function.  In one-dimension, $\nabla^2 = \frac {\partial^2} {\partial x^2}$.    $S$ is given by
\begin{equation}\label{eq:spinteract}
S = \left\{\begin{array}{cl}
\phantom{-} \frac{\hbar^{2}}{4} & \quad\mbox{(triplet)} \\
-\frac{3\hbar^{2}}{4} & \quad\mbox{(singlet)} .
\end{array}\right.
\end{equation}
Note that the parameters that we obtain during numerical optimization may be unphysical in the sense the interaction energies that result can be much larger than is typically observed between electrons; but, we stress that our goal is to understand the extremes of what is consistent with the Schr\"{o}dinger equation.  As discussed later, the resulting universal properties are unchanged if the parameters are constrained to a physically-reasonable range.

The piece-wise potential energy function is the most general one-dimensional potential energy function insofar as the pieces are small compared with the size of the smallest features of the system.  The form of the interaction potential, on the other hand, is precisely the one-dimensional coulomb interaction.  Thus, {\em our model accounts for any system with interacting point charges}, of which a two-electron one-dimensional molecule is a subset.

For most of the experiments reported here we use
$C_{1} = 0.1$, $C_{2} = 0.5$, $C_{3} = 0.1$.  These are
\emph{ad hoc} choices.  We found that our results are not too
sensitive to the choice of constants.
However, if $V_{21}$ is made too large relative to
$V_{22}$, the electrons are separated into two separate wells and we get small
values of $\beta_{int}$.  In some experiments reported below, we
allowed the constants to vary.  We found that the spin interaction
term $V_{22}$ is much more important for making the intrinsic hyperpolarizability larger than is the electrostatic repulsion term $V_{21}$.

Subspaces of symmetric and antisymmetric functions are invariant under
any Hamiltonian of the form Given by Equation \ref{eq:schro_yesinteract}.
Therefore we can consider the symmetric and
antisymmetric eigenvalue problems separately.  For the symmetric
spatial wavefunctions ($\Psi(x_{2},x_{1}) = \Psi(x_{1},x_{2})$) the spin part is a
singlet, while for the antisymmetric spatial wavefunctions
($\Psi(x_{2},x_{1}) = -\Psi(x_{1},x_{2})$) the spin part is a triplet.

We approximate Equation \ref{eq:schro_yesinteract} on a $20 \times 20$
grid of quadratic serendipity finite elements \cite{zienk05.01}.   On
the symmetric subspace the number of degrees of freedom is cut nearly
in half by the restriction $\Psi(x_{2},x_{1})=\Psi(x_{1},x_{2})$.
On the antisymmetric
subspace it is more than halved.
On each of the subspaces we use the implicitly restarted
Arnoldi method \cite{soren92.01} to compute the wave functions and
energy levels.

In all of our computations, the ground state is a
symmetric function (spin singlet), so the value of $\beta$ depends
only on the states that are symmetric in space.  To see this, we
consider the definition of the transition moments $x_{0m}$, which
appear as factors in every term of the SOS or dipole-free expression for $\beta$,
$$x_{0m} = \int_{0}^{L}\int_{0}^{L}(t_{1} +
t_{2})\Psi_{0}(t_{1},t_{2})\Psi_{m}(t_{1},t_{2}) \, dt_{1} \, dt_{2}.$$
If $\Psi_{0}$ is symmetric and $\Psi_{m}$ is antisymmetric, then the
integrand is antisymmetric and the integral $x_{0m}$ is zero.  Thus
this term contributes nothing to $\beta$.

In addition to the computations of $\beta_{int}$, we compute the
matrix\cite{zhou06.01,kuzyk06.01}
\begin{equation}
\tau_{mp}^{(N)} = \delta_{m,p} - \frac{1}{2}\sum_{n=0}^{N} \left(
  E_{nm} + E_{np}\right) x_{mn} \,x_{np} , \label{tau}
\end{equation}
where $N$ is the number of states used.
Each matrix element of $\tau^{(N)}$, indexed by $m$ and $p$, is a measure of how well the $(m,p)$ sum rule is obeyed when truncated to $N$ states.  If the sum rules are exactly obeyed, $\tau_{mp}^{(\infty)}=0$ for all $m$ and $p$.  We note that if the sum rules are truncated to an N-state model, the sum rules indexed by a large value of $m$ or $p$ (i.e. $m,p \sim N$) may be disobeyed even when the position matrix elements and energies are exact.  We have found that the values of $\tau_{mp}^{(N)}$ are small for exact wavefunctions when $m<N/2$ and $p<N/2$.  So, when evaluating the $\tau$ matrix to test our calculations, we consider only the components $\tau_{m\leq N/2,p\leq N/2}^{(N)}$.

Since
the hyperpolarizability depends critically on the transition dipole
moment from the ground state to the excited states, we use the value
of $\tau_{00}^{(30)}$ as one important test of the accuracy of the
calculated wavefunctions.  Additionally, we use the standard deviation
of $\tau^{(N)}$,
\begin{equation}
\Delta \tau^{(N)} = \frac{\sqrt{ \sum_{m=0}^{N/2} \sum_{p=0}^{N/2} \left( \tau_{mp}^{(N)} \right)^2 }} {N/2} , \label{Dtau}
\end{equation}
which quantifies, on average, how well the sum rules are obeyed in aggregate, making $\Delta \tau^{(N)}$ a broader test of the accuracy of a large set of wavefunctions.

We mentioned earlier that we did our computations on a $20 \times 20$
grid of finite elements.  As a check on whether this mesh is fine
enough, we repeated our final computations on a $100 \times 100$ grid
and found that we got the same results.

Our code is written in MATLAB.

\section{Results and Discussions}

\begin{widetext}

\begin{table}
\caption{Summary of results for two interacting electrons -- with $C_{1} = 0.1$, $C_{2} = 0.5$, and $C_{3} = 0.1$ -- using different starting
    potentials.  $\beta_s$ is the intrinsic hyperpolarizability
    of the starting
    potential while $\beta_{int}$ is the value after optimization.  The
    transition moments and energies are in dimensionless units.  To
    convert to specific units, consider an example where $x_{nm}$ is
    interpreted to be in units of angstroms.  The energies would then
    be determined by multiplying all values of $E_{n0}$ by
    $\hbar^2/ma^2$, with $a=10^{-10} \, m$ (1\,{\AA}).  In this
    case, the energy is in units of $1.2 \times 10^{-18} \, J$ or
    about $7.6 eV$.\label{tab:Vsummary}}
\begin{tabular}{c c c c c c c c c c c }
  \hline
  Function & $\beta_{S}$ & $\beta_{int}$
  & $\tau_{00}^{(80)}$ & $\Delta \tau^{(80)}$ & $E_{10}$ & $E_{20}$ &
  $E_{10}/E_{20}$ & $x_{00}$ & $x_{10}$ & $\frac {x_{10}} {x_{10}^{max}}$ \\
 $V(x)$ &  & & & &  &  \\
  \hline

0 & 0 & 0.7068 & 0.0027 & 0.0704 & 0.0235 &  0.0488
& 0.4816 & 11.2267 & -5.1493 & 0.7888 \\

$x$ & 0.6319 & 0.7063 & 0.0044 & 0.0574 & 0.0253 &
 0.0528 & 0.4787 &  -11.8570 & -4.9505 & 0.7872 \\

$x^{2}$ & 0.5523 & 0.7064 & 0.0035 & 0.0589 & 0.0249
& 0.0520 & 0.4799 & -11.7233 & -4.9913 &  0.7883  \\

$\sqrt{x}$ & 0.6648 & 0.7063 & 0.0042 & 0.0599 &
0.0259 & 0.0540 & 0.4808 & -12.0313 & -4.8971 & 0.7888 \\

$x + \cos(x)$ & 0.0311 & 0.7021 & 0.0044 & 0.0631 &
0.0381 & 0.0792 & 0.4808 & -11.0082 & -4.0336 & 0.7869 \\

$\tanh(x)$ & 0.0018 & 0.7068 & 0.0024 & 0.0601 &
0.0234 & 0.0488 & 0.4800 & -11.2016 & -5.1530 & 0.7886 \\

random     & 0.1066 & 0.7022 & 0.0080 & 0.0770 &
0.0395 & 0.0832 & 0.4749 &  14.6972 & 3.9642 & 0.7882 \\
\hline
\end{tabular}
\end{table}

\end{widetext}

We use various analytical functions as starting potentials, and then apply the numerical optimization technique as described above to determine the optimal potential function for each.  The tables that follow give the hyperpolarizability of the starting potential, $\beta_S$ followed by the hyperpolarizabilities of the optimized potentials.  So, for example, when $V=0$, we see that $\beta_S = 0$; but, the optimized potentials are all nonzero.
 
Because the Nelder-Mead algorithm delivers only a local maximum,
the final optimized potential can vary, depending on the choice of the
starting potential.  Table \ref{tab:Vsummary} summarizes
the result of optimizing $\beta_{int}$ for two interacting electrons  with $C_{1} = 0.1$, $C_{2} = 0.5$, and $C_{3} = 0.1$ when starting from six different potentials, which are the same as in previous work\cite{zhou07.02} except that they have been re-scaled for convenience.  Each starting potential (except $V(x)=0$) was scaled so that $V(L) = \hbar^{2}/10$.  Also, when we state that our starting potential is $V(x) = x^2$, for example, we mean
that it is the piecewise linear interpolant of $x^2$.

It is important to stress that our approach is based on invariance under simple scaling,\cite{kuzyk10.01} so that the magnitudes of the energies, dipole matrix elements, and hyperpolarizabilities are unimportant to the analysis.  Rather, ratios of the observables, such as $E_{10}/E_{20}$, $x_{10} / x_{10}^{max}$ and $\beta/\beta_{max}$ are the key intensive quantities.  As such, any row of Table \ref{tab:Vsummary} can be transformed by simple scaling to arbitrarily adjust all the quantities, but leaving the intensive parameters unaffected.  It is this property of simple scaling that makes possible a comparison of a broad range of molecular shapes and sizes using the intrinsic hyperpolarizability as the metric.

The values of $\beta_{int}$ reported in the table were computed using
$\beta_{int} = \beta_{NP}/\beta_{MAX}$.  When we checked the
computation using $\beta_{SOS}$ in place of $\beta_{NP}$, we got the
same values to at least four decimal places, thus, we choose to use $\beta_{NP}$ in all of our optimization calculations.  When we used
$\beta_{DF}$, we had agreement to within a fraction of a percent.  In the last column of Table \ref{tab:Vsummary}, $x_{10}^{max}$ is the magnitude of the fundamental limit of the position matrix element
$x_{10}$ for a two-electron system, and is given by,
\begin{equation}
x_{10}^{max} = \frac {\hbar} {\sqrt{m E_{10}}} . \label{x10MAX}
\end{equation}

As was found for the one-electron case,\cite{zhou07.02,kuzyk10.01} a quantum system with a hyperpolarizability near the fundamental limit shares certain universal properties.  For example, all optimized values of $\beta_{int}$ are just over 0.7, for which the energy ratio $E_{10}/E_{20}$ is just under 0.5 and the intrinsic transition moment $\frac {x_{10}} {x_{10}^{max}}$ is about 0.79.  Note that if electron interaction as characterized by the $C$ parameters is decreased smoothly to zero, the optimized hyperpolarizability approaches 0.5, as is predicted analytically for the non-interacting limit.  When electron interactions are made very large, the optimized intrinsic hyperpolarizability does not increase beyond $\beta_{int} = 0.708$.
\begin{widetext}

\begin{table}\caption{Summary of results for the hyperpolarizability computed from the
    subspace of antisymmetric functions (triplet states).
    $\beta_s$ is the intrinsic hyperpolarizability
    of the starting
    potential while $\beta_{int}$ is the value after optimization.  The
    transition moments and energies are in the same dimensionless units as in Table \ref{tab:Vsummary}.\label{tab:Vsummary_anti}}
\begin{tabular}{c c c c c c c c c c c}
  \hline
  Function & $\beta_{S}$ & $\beta_{int}$
  & $\tau_{00}^{(30)}$ & $\Delta \tau^{(30)}$ & $E_{10}$ & $E_{20}$ &
  $E_{10}/E_{20}$ & $x_{00}$ & $x_{10}$ & $\frac {x_{10}} {x_{10}^{max}}$ \\
 $V(x)$ &  & & & &  &  \\
  \hline

0 & 0 & 0.6692 & 0.0014 & 0.0546 &  0.0528 & 0.1153 & 0.4582 &
5.7865 & 3.4477 & 0.7925 \\

$x$ & 0.0636 & 0.6690 & 0.0056 & 0.1472 & 0.0813 &
0.1817 & 0.4472 & 1.5841 & 2.7723  & 0.7902 \\

$x^{2}$ & 0.0057 & 0.6670 & 0.0025 & 0.1181 & 0.0650 &
0.1426 & 0.4556 & 2.4667 & 3.0848 &  0.7863 \\

$\sqrt{x}$ & 0.1047 & 0.6682 & 0.0001 & 0.0999 & 0.0713 &
0.1582 & 0.4507 & 2.0296 & 2.9681 & 0.7926 \\

$x + \cos(x)$ & 0.2113 & 0.6676 & 0.0023 & 0.0614 & 0.0782 &
0.1664 & 0.4698 & 4.4549 & 2.8294 & 0.7912 \\

$\tanh(x)$ & 0 & 0.6691 & 0.0032 & 0.0570 & 0.0538 & 0.1174 &
0.4586 & 7.4199 & 3.4047 & 0.7900 \\
\hline
\end{tabular}
\end{table}

\end{widetext}

While we have focused mainly on well-defined potential energy functions, it is also possible to apply Monte Carlo methods to randomly pick the shape of the potential.  In the first simulation, we picked the random numbers to be normally distributed with mean zero and standard deviation 1.  The optimization algorithm got trapped in the parameter space where $\beta_{int}$ was small.  However, for a set of random numbers with the standard deviation reduced to 0.1, the simulation gave $\beta_{int} \approx 0.7$.  The resulting parameters are shown in the last row of Table \ref{tab:Vsummary}.  Once again, the same universal parameters are observed as for the other optimized potentials.

It is not likely that the universal properties that we observe are coincidental or artifacts of the calculation given independent calculations using widely different approaches.  In addition to the large number of one-dimensional potentials for one and two electrons, for which these same universal properties result,\cite{zhou06.01,zhou07.02}  {\em the same universal properties are found} when allowing point nuclei to move in a plane to form a molecule,\cite{kuzyk06.02} applying electromagnetic fields to two-dimensional molecules,\cite{watkins09.01} and using constrained potentials with properties more similar to real molecules and including continuum states.\cite{szafr10.01} Monte Carlo calculations, which probe a much larger parameter space than the former examples are also observed to share some universal properties, though they are less restrictive than in the cases where a potential energy function and vector potential is used.

In all simulations, the energies of the triplet states were tracked and found to be of higher energy than the ground state singlet state.  Since the triplet states are orthogonal to the space of singlet states, no triplet states contribute to the results.  To investigate the hyperpolarizability due to the triplet states, we used the lowest energy triplet state to calculate the dipole moment from which the hyperpolarizability is determined using the nonperturbative method.  We also used the SOS expressions with the lowest energy triplet state as the ground state to calculate the hyperpolarizability.  Thus, these calculations only consider transitions between triplet states. Table \ref{tab:Vsummary_anti} summarizes the results.

In all cases, the optimized intrinsic hyperpolarizability for the triplet subspace is just short of 0.67, which is lower than the value of 0.71 found in the singlet subspace. Similarly, the other parameters, such as the energy ratio and normalized transition moment, are close but not identical to the singlet values.  Thus, we conclude that the qualities of the optimized system using only triplet states is similar to those of the singlet subspace.  However, when using the true ground state of the system, the same universal properties are observed for two interacting electrons as is observed for a single electron.

\begin{figure}
\includegraphics[width=9cm]{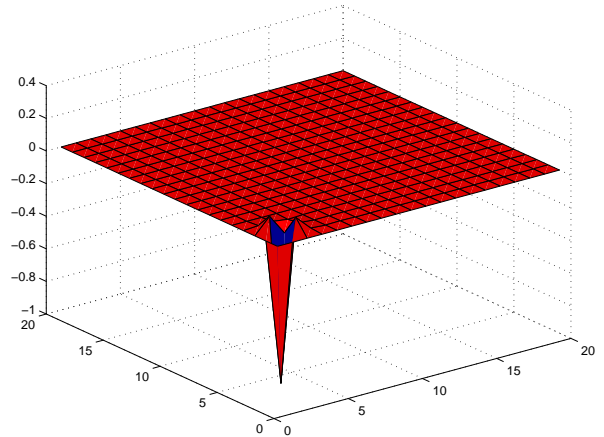}
\caption{Surface plot of $\beta_{nm}$ that is representative of all the optimized hyperpolarizabilities in Table \ref{tab:Vsummary}.}
\label{fig:V=0-betanm}
\end{figure}
We define $\beta_{nm}$ as the contribution from states $n$ and $m$ to the SOS sum,\cite{zhou07.02}
\begin{equation}\label{betanm}
\beta_{nm} = \frac{x_{0n} \overline{x}_{nm} x_{m0}}{E_{n0} E_{m0}},
\end{equation}
where $\overline{x}_{nm} \equiv x_{nm} - x_{00} \delta_{n,m}$.  The
hyperpolarizability is calculated by summing $\beta_{nm}$ over all
excited states of the system.  Figure \ref{fig:V=0-betanm} shows a
plot of $\beta_{nm}$ that is representative for all the potentials
plotted in Figure \ref{tab:Vsummary}.  Typical values are  $\beta_{11}
= - 0.9$,  $\beta_{21} = \beta_{12} = 0.09$, and all other matrix
elements are at least 10 times smaller.  Thus, the non-interacting
electron potentials that optimize $\beta_{int}$ lead to
hyperpolarizabilities that obey the three-level
ansatz,\cite{zhou06.01} that is, a system with $\beta$ near the
fundamental limit is described by a three-level model.

The three-level ansatz has been shown to be universally obeyed in systems described by a one-dimensional potential,\cite{zhou07.02} point nuclei in a plane,\cite{kuzyk06.02} and for one-dimensional monte-carlo simulations.\cite{kuzyk08.01} In the case of arbitrarily applied electric fields to point nuclei in two-dimensions, the optimized diagonal component of the hyperpolarizability obeys the three-level ansatz.  However, optimized $\beta_{xxy}$ yields $\beta_{int} = \beta_{xxy}/\beta_{xxx}^{max} \approx 0.9$ and the intrinsic transition moment is $0.51$.  In this case, the hyperpolarizability is dominated by two excited states; but, an additional pair of excited states contribute about 25\%.

\begin{figure}
\includegraphics[width=9cm]{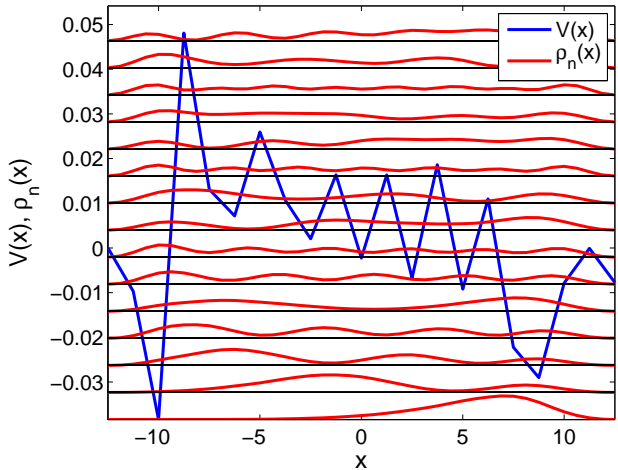}
\caption{Optimized potential energy function and electron density for the lowest-energy
  15 states of a 30-state model. Starting potential is $V(x)=0$.}
\label{fig:V=0-potential}
\end{figure}

In further experiments, starting with the potential depicted in Figure
\ref{fig:V=0-potential}, we attempted to increase $\beta_{int}$
even further by varying the constants $C_{1}$, $C_{2}$, and $C_{3}$
appearing in the interaction terms (Equation \ref{eq:v21} and Equation \ref{eq:v22}).  These constants are all constrained to be positive.  We achieved only a modest increase from $\beta_{int} = 0.7068$ to
$\beta_{int} = 0.7080$, still no better than the largest intrinsic hyperpolarizability attained by varying the potential energy function;\cite{zhou07.02} but the largest found in our simulations.  The final values of $C_{1}$, $C_{2}$, and $C_{3}$ were 0.0034, 0.8787, and 0.4942.  The strength of the
electrostatic repulsion term is governed by $C_{1}$, which decreased
from 0.1 to 0.0034 during the optimization.  Similar results were seen
in other experiments of this type.  Thus the electrostatic repulsion
term seems not to be important for maximizing $\beta_{int}$.  The
strength of the spin interaction term is governed by $C_{2}$, which
increased from 0.5 to 0.8787 during optimization.  This suggests
that the spin interaction term is important for maximizing
$\beta_{int}$.

\begin{widetext}

\begin{table}\caption{Parameters resulting from optimizing the intrinsic hyperpolarizability starting with
   the potentials shown and varying the strengths of the electron interaction terms.  For the $V(x)=0$ starting potential, both the electron interaction parameters and potential function were varied.
    $\beta_s$ is the intrinsic hyperpolarizability of the starting
    potential while $\beta_{int}$ is the value after optimization.
    \label{tab:Vsummary2}}
\begin{tabular}{c c c c c c c c c c c c}
  \hline
  Function & $\beta_{S}$ & $\beta_{int}$
  & $\tau_{00}^{(80)}$ & $\Delta \tau^{(80)}$ & $E_{10}$ & $E_{20}$ &
  $E_{10}/E_{20}$ & $x_{10}$ & $\frac {x_{10}} {x_{10}^{max}}$ \\
 $V(x)$ &  & & & &  &  \\
  \hline\hline

 Figure \ref{fig:V=0-potential}   & 0.7068 & 0.7080 &  0.0018 & 0.0627 & 0.0235 & 0.0490 & 0.4803 & 5.1265  & 0.7865 \\

$x$ & 0.6319 & 0.6663 & 0.0069  & 0.0149  & 0.0442  & 0.0804 & 0.5494  & -4.1252  & 0.8670
\\

$0$ &  0  & 0.7074 & 0.0010 & 0.0549 & 0.0249 & 0.0516 & 0.4827 & -4.9887 & 0.7870
\\ \hline

\end{tabular}
\end{table}

\end{widetext}

\begin{table}\caption{Electron interaction parameters resulting from optimizing the intrinsic hyperpolarizability starting with the potentials shown and varying the strengths of the electron interaction terms.  For the $V(x)=0$ starting potential, both the electron interaction parameters and potential function were varied.  $\beta_s$ is the intrinsic hyperpolarizability of the starting potential while $\beta_{int}$ is the value after optimization.
    \label{tab:Csummary2}}
\begin{tabular}{c c c c c c}
  \hline
  Function & $\beta_{S}$ & $\beta_{int}$ & $C_1$ & $C_2$ & $C_3$ \\
 $V(x)$ &  & & & &\\
  \hline\hline

 Figure \ref{fig:V=0-potential}   & 0.7068 & 0.7080 & 0.0034 & 0.8787 & 0.4942 \\

$x$ & 0.6319 & 0.6663 & 0.000087  & 14.75 & 0.0612\\

$0$ &  0  & 0.7074 & 0.0933 & 0.5283 &  0.1033 \\
\hline

\end{tabular}
\end{table}

In another experiment we used the potential $V(x)=x$ and started with
$C_{1} = 0.1$, $C_{2}=0.5$, and $C_{3} = 0.1$.  Holding $V(x)$ fixed,
we tried to increase $\beta_{int}$ by varying  $C_{1}$, $C_{2}$, and
$C_{3}$.  With this potential we already have a rather good
$\beta_{int} = 0.6319$ to begin with.  By varying the constants
$C_{i}$, we were only able to realize a modest increase in
$\beta_{int}$ to about 0.6663, which was obtained with $C_{1} =
0.000087$, $C_{2} = 14.75$, and $C_{3} = 0.0612$.  With such a small
$C_{1}$, the electrostatic repulsion term is essentially zero.  The
large value of $C_{2}$ means that the spin interaction term is very
strong, and the small value of $C_{3}$ means that the spin interaction operates only over a very short range.  We conclude that varying the electron interaction potential alone does not have a large effect on improving the intrinsic hyperpolarizability.

In another experiment we started with the potential $V(x)=0$ and the
initial constants $C_{1}= 0.1$, $C_{2} = 0.5$, and $C_{3} = 0.1$.  When allowing both the potential energy function and electron interaction parameters to vary, we obtained $\beta_{int} = 0.7074$.  The final
values of the electron interaction parameters were $C_{1} = 0.0933$, $C_{2} = 0.5283$, and $C_{3} = 0.1033$,
so they changed little from their initial values.

Table \ref{tab:Vsummary2} summarizes some of the parameters found for the optimized intrinsic hyperpolarizabilities when electron interactions are present.  In the two cases where the intrinsic hyperpolarizability is greater than 0.7, the same universal values are observed; i.e. the energy ratios and the intrinsic transition moments are almost identical to the ones in Table \ref{tab:Vsummary}.  Furthermore, $\beta_{nm}$ for all three optimized potentials are found to be dominated by only one pair of excited states.  Thus, the three-level ansatz is again obeyed when electron interactions are important.

Based on these observations, we conclude that when the diagonal component of the hyperpolarizability is optimized, all systems - including ones with two interacting electrons - share the same universal properties; and, the three-level ansatz holds.  This observation suggests that when investigating the properties of a quantum system, the numerical optimization of the simplest models can be used to identify universal properties.  This paradigm is in stark contrast to studies that seek to understand the properties of a specific system,\cite{fu07.01,fu07.02} and is more akin to the theoretical studies by Meyers et al, who studied the effect of an electric field on polyenes\cite{meyer94.01} which confirmed the concept of bond length alternation.\cite{marde91.01,marde93.01} However, our approach of optimizing the nonlinear response using the potential energy function is even more general.  The only constraint we apply is that the sum rules are obeyed, which demands consistency with quantum mechanics but is otherwise general.

Arguably, the most important result of universal property studies is that there appear to be a very large number of ways to make a system with a nonlinear-optical response near the fundamental limit with $\beta_{int} = 0.7$.\cite{kuzyk10.01}  Szafruga and coworkers performed numerical optimization studies under the constraint that the potential energy varies by an amount that is typical in a molecule, as well as extending the computational space beyond the potential well to approximate continuum states.\cite{szafr10.01}  It was found that with those restrictions the same universal properties were observed, and $\beta_{int} \approx 0.708$.  Therefore, it appears that it should be possible to make many classes of yet-to-be-found molecules that are at the fundamental limit.

Some of the parameters shown in Table \ref{tab:Csummary2} yield electron interaction energies that are larger than typically observed.  However, even in the strong interaction limit, the effect on the optimized hyperpolarizability is not large.  Thus, we conclude that when electron interactions are included, the universal properties remain the same and that electron interactions may often be unimportant when studying universal properties of optimized systems.  This is not to say that electron interactions are unimportant when calculating the properties of a specific molecule.  When we limited the parameter space in the present studies in the way described in Szafruga\cite{szafr10.01}, the observed universal properties with electron interactions remained the same.

\begin{figure}
\includegraphics[width=9cm]{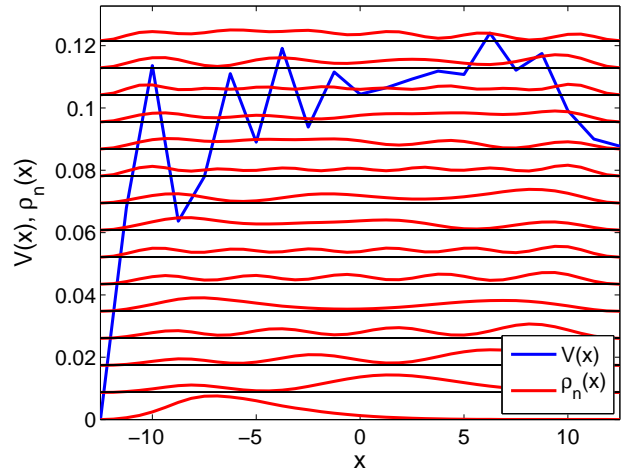}
\caption{Optimized potential energy function and electron density for
  15 states. Starting potential is $V(x)= \tanh(x)$.}
\label{fig:tanh-potential}
\end{figure}

Finally, we compare the charge densities that result for the optimized potential energy function with two electrons with prior results with one electron.  Figure \ref{fig:V=0-potential} and Figure \ref{fig:tanh-potential} show the optimized potential energy functions for starting potentials $V(x) = 0$ and $V(x) = \tanh(x)$.  Also shown is the electron density for the lowest-energy fifteen states of a thirty -state model.  Note that the numerical labels on the y axes refer only to the potential energy functions and all electron densities are positive.  This is a common observation that very different potential energy functions that lead to very different wavefunctions always share the same universal properties.  Thus, there is a unifying underlying structure that is shared by these optimized systems that are not apparent simply from the electron densities.

In past studies,\cite{zhou07.01,zhou07.02} it was found that two classes of wavefunction lead to optimized hyperpolarizability: (1) Localized charge density resulting from randomly oscillating potentials, reminiscent of Anderson Localization for the $\tanh(x)$ starting potential, and (2) delocalized wavefunctions.  In both cases, two excited states dominated in their contribution to the hyperpolarizability.  In the present studies, there is a hint of localization in the ground and first excited states, but the others are delocalized.  However, this result may be due to the fact that the present calculation used piecewise continuous potentials rather than splines, which were used in previous studies.

\begin{figure}
\includegraphics[width=9cm]{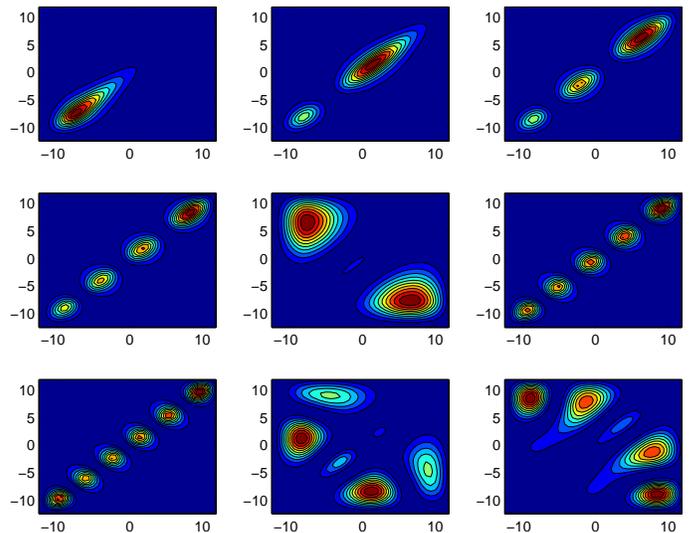}
\caption{Contour plots of electron density functions for the nine
  lowest energy levels (left to right, top to bottom) for the optimized potential with starting potential $V=\tanh x$.}
\label{fig:tanh-densities}
\end{figure}

To study localization, it is more instructive to consider the full two-electron wavefunctions.  Figure \ref{fig:tanh-densities} shows the electron density $\left| \psi_n (x_1,x_2) \right|^2$ as a function of $x_1$ and $x_2$ for an optimized quantum system using the starting potential $V(x) = \tanh(x)$.  The electrons are correlated in the four lowest energy eigenstates.  Many of the higher states show that the two electrons are localized in different spatial regions.  This is reminiscent of the one-dimensional one-electron potential, where the wave functions do not overlap.  Note that the electron densities appear similar in the excited states of all optimized systems.

\begin{figure}
\includegraphics[width=9cm]{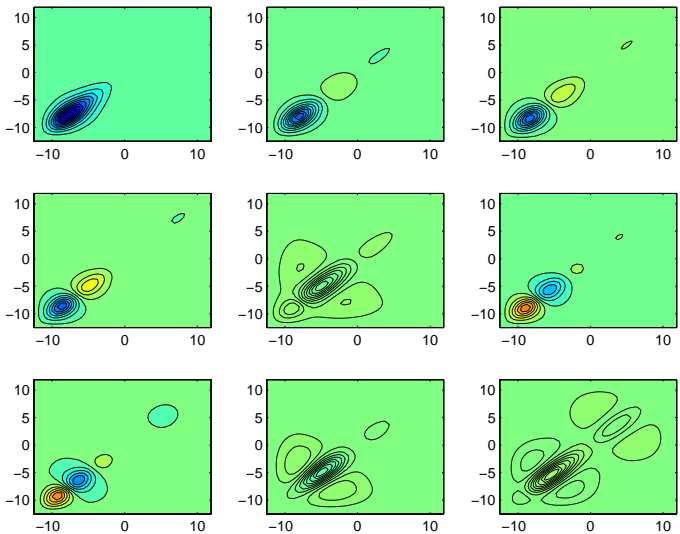}
\caption{Contour plots of transition dipole moment amplitude  $\psi_n (x_1,x_2) \left( x_1 + x_2 \right) \psi_0 (x_1,x_2)$ with $n = 1$ to $9$ (left to right and top to bottom) for the optimized potential energy function with starting potential $V=\tanh x$.}
\label{fig:V=tanhdipoleContour}
\end{figure}

The transition dipole moment amplitude from the ground state,  $\psi_n (x_1,x_2) \left( x_1 + x_2 \right) \psi_0 (x_1,x_2)$, is shown in Figure \ref{fig:V=tanhdipoleContour}.  The only transition dipole moment densities that are large are for transitions from the ground state to the first and second excited states.  For the higher states, the transition moment amplitudes are either small; or, large positive regions are balanced by small negative regions. Thus, the three-level ansatz is obeyed either through minimal overlap between the different eigenfunctions, or cancelation between positive and negative probability amplitude.

One might ask how the potential energy function can be manipulated to make a real system with an optimized nonlinear optical response.  There are many artificial structures, such as nanoparticles, quantum wires, and multiple quantum wells in which there is some control over the structure and the resulting potential profile.  For example, nanowires can be used as building blocks for nanoscale electronic and optoelectronic devices,\cite{duan2001indium} and multiple quantum wells can be made, for example, with a band edge that can be used to make a system with a large quadratic electroabsorption coefficient.\cite{miller1984band} Recently, Hall and coworkers used graded barriers to control the optical properties of ZnO/ZnMgO quantum wells with an intrinsic internal electric field.\cite{hall2010using} Clearly, there are many alternatives to molecules that can be used to deliberately engineer a potential profile.

Wang and coworkers have shown that the potential energy profile of a molecule can be approximated by using a linear combination of atomic potentials.\cite{wangm06.01} Thus, the results of our work may be able to be connected to real molecules using such an approach.  We stress that optimization of the potential energy function is a more general approach than other numerical optimization techniques, with the exception of the Monte Carlo techniques,\cite{kuzyk08.01,shafei10.01} because potentials apply to the broader set of systems of which molecules are a small subset.  Our present work shows that the observed universal properties are not restricted to just one electron systems.

It is no simple matter to determine the nature of the interaction between electrons that leads to the $N^{3/2}$ scaling law.  The fact that the fundamental limit of the hyperpolarizability scales as $N^{3/2}$ follows from the sum rules and the limits they set for the oscillator strengths.  If the full oscillator strength resides in one state, the modulus squared of the transition moment to that state from the ground state is proportional to the number of electrons, thus, the transition moments scale as the square root of the number of electrons.  Since the hyperpolarizability scales as the third power of the transition moment, it scales as $N^{3/2}$.  Note that the sum over the dipole terms of the form $\Delta x_{n} \left| x_{n0} \right|^2$ can be re-expressed using the sum rules in terms of sums of the products of the form $x_{0n} x_{nm} x_{m0}$.\cite{kuzyk05.02}  Thus, while counterintuitive, the dipole terms acting together will also scale as $N^{3/2}$.

In the absence of interactions, the optical properties are obviously additive over the molecular units.  When allowed to interact, the largest power law allowed by the fundamental limits is $N^{3/2}$, and the optimization algorithm finds this local maximum. The fact that the parameters that characterize the interaction show no universal behavior (see Table \ref{tab:Csummary2}) implies that there is no unique interaction that leads to the optimal response.  Thus, the question about the nature of the interactions that leads to optimized scaling does not have a unique answer.

In closing, we note that the focus of our present work is on the {\em off-resonant } hyperpolarizability. Since the fundamental limit of the dispersion of the hyperpolarizability has also been calculated,\cite{kuzyk06.03} it would be a straightforward matter to calculate the potential energy function that is optimized at any set of wavelengths.  Such calculations would be useful in determining the optical material properties required for a particular device.  Since there are an infinite number of combinations of wavelengths, we have limited our present studies to the off-resonance regime.

\section{Conclusions}

Rather than calculating the nonlinear-optical response of large numbers of complex molecules and searching for correlations between structure and nonlinear response, our work focuses on developing general principles that broadly apply to all quantum systems; and, endeavors to identify universal properties that can be applied to build an understanding of what leads to a large nonlinear-optical susceptibility.

The present work seeks to determine whether or not electron interactions and Fermi correlations change the universal behavior of an optimized system as is observed for one electron systems.  We have found that the universal behavior is indeed the same.  The intrinsic transition moment, energy ratio of the first two dominant excited states, and the intrinsic hyperpolarizabilities for an optimized system are all the same for a broad range of starting potentials.  Also, the three-level ansatz is again found to be obeyed.  Since electron interactions appear to have a small effect on the conclusions derived from numerical optimization, simple one-electron models may be enough to study the broad principles underlying light-matter interactions.

Optimization studies are inherently tricky because one can never be certain whether or not a global maximum has been found.  However, as we continue to generate more numerical data from a broader range of starting potentials, molecular symmetries, externally-applied electromagnetic fields, and electron interactions, we continue to observe the same universal properties - leading to the tentative conclusion that perhaps the data as a whole is representative of the properties of the true global maxima.  Similarly, with only a handful of exceptions, experimental work that focuses on the development of new molecules finds that the hyperpolarizabilities are far short of the fundamental limit.  The broad principles that we seek to develop aims to bridge the gap between the limits and real molecular systems.  Clearly, altogether new paradigms will need to be developed to attain what is possible.

{\bf Acknowledgements: } MGK thanks the National Science Foundation (ECCS-0756936) and Wright Paterson Air Force Base for generously supporting this work.

%\bibliography{\bibs}

\begin{thebibliography}{56}%
\makeatletter
\providecommand \@ifxundefined [1]{%
 \@ifx{#1\undefined}
}%
\providecommand \@ifnum [1]{%
 \ifnum #1\expandafter \@firstoftwo
 \else \expandafter \@secondoftwo
 \fi
}%
\providecommand \@ifx [1]{%
 \ifx #1\expandafter \@firstoftwo
 \else \expandafter \@secondoftwo
 \fi
}%
\providecommand \natexlab [1]{#1}%
\providecommand \enquote  [1]{``#1''}%
\providecommand \bibnamefont  [1]{#1}%
\providecommand \bibfnamefont [1]{#1}%
\providecommand \citenamefont [1]{#1}%
\providecommand \href@noop [0]{\@secondoftwo}%
\providecommand \href [0]{\begingroup \@sanitize@url \@href}%
\providecommand \@href[1]{\@@startlink{#1}\@@href}%
\providecommand \@@href[1]{\endgroup#1\@@endlink}%
\providecommand \@sanitize@url [0]{\catcode `\\12\catcode `\$12\catcode
  `\&12\catcode `\#12\catcode `\^12\catcode `\_12\catcode `\%12\relax}%
\providecommand \@@startlink[1]{}%
\providecommand \@@endlink[0]{}%
\providecommand \url  [0]{\begingroup\@sanitize@url \@url }%
\providecommand \@url [1]{\endgroup\@href {#1}{\urlprefix }}%
\providecommand \urlprefix  [0]{URL }%
\providecommand \Eprint [0]{\href }%
\@ifxundefined \urlstyle {%
  \providecommand \doi  [0]{\begingroup \@sanitize@url \@doi}%
  \providecommand \@doi [1]{\endgroup \@@startlink {\doibase
  #1}doi:\discretionary {}{}{}#1\@@endlink }%
}{%
  \providecommand \doi  [0]{doi:\discretionary{}{}{}\begingroup
  \urlstyle{rm}\Url }%
}%
\providecommand \doibase [0]{http://dx.doi.org/}%
\providecommand \Doi [0]{\begingroup \@sanitize@url \@Doi }%
\providecommand \@Doi  [1]{\endgroup\@@startlink{\doibase#1}\@@Doi}%
\providecommand \@@Doi [1]{#1\@@endlink}%
\providecommand \selectlanguage [0]{\@gobble}%
\providecommand \bibinfo  [0]{\@secondoftwo}%
\providecommand \bibfield  [0]{\@secondoftwo}%
\providecommand \translation [1]{[#1]}%
\providecommand \BibitemOpen [0]{}%
\providecommand \bibitemStop [0]{}%
\providecommand \bibitemNoStop [0]{.\EOS\space}%
\providecommand \EOS [0]{\spacefactor3000\relax}%
\providecommand \BibitemShut  [1]{\csname bibitem#1\endcsname}%
%</preamble>
\bibitem [{\citenamefont {Chen}\ \emph {et~al.}(2004)\citenamefont {Chen},
  \citenamefont {Kuang}, \citenamefont {Wang},\ and\ \citenamefont
  {Sargent}}]{wang04.01}%
  \BibitemOpen
  \bibfield  {author} {\bibinfo {author} {\bibfnamefont {Q.~Y.}\ \bibnamefont
  {Chen}}, \bibinfo {author} {\bibfnamefont {L.}~\bibnamefont {Kuang}},
  \bibinfo {author} {\bibfnamefont {Z.~Y.}\ \bibnamefont {Wang}}, \ and\
  \bibinfo {author} {\bibfnamefont {E.~H.}\ \bibnamefont {Sargent}},\
  }\href@noop {} {\bibfield  {journal} {\bibinfo  {journal} {Nano. Lett.},\
  }\textbf {\bibinfo {volume} {4}},\ \bibinfo {pages} {1673} (\bibinfo {year}
  {2004})}\BibitemShut {NoStop}%
\bibitem [{\citenamefont {Cumpston}\ \emph {et~al.}(1999)\citenamefont
  {Cumpston}, \citenamefont {Ananthavel}, \citenamefont {Barlow}, \citenamefont
  {Dyer}, \citenamefont {Ehrlich}, \citenamefont {Erskine}, \citenamefont
  {Heikal}, \citenamefont {Kuebler}, \citenamefont {Lee}, \citenamefont
  {McCord-Maughon}, \citenamefont {Qin}, \citenamefont {Rockel}, \citenamefont
  {Rumi}, \citenamefont {Wu}, \citenamefont {Marder},\ and\ \citenamefont
  {Perry}}]{cumps99.01}%
  \BibitemOpen
  \bibfield  {author} {\bibinfo {author} {\bibfnamefont {B.~H.}\ \bibnamefont
  {Cumpston}}, \bibinfo {author} {\bibfnamefont {S.~P.}\ \bibnamefont
  {Ananthavel}}, \bibinfo {author} {\bibfnamefont {S.}~\bibnamefont {Barlow}},
  \bibinfo {author} {\bibfnamefont {D.~L.}\ \bibnamefont {Dyer}}, \bibinfo
  {author} {\bibfnamefont {J.~E.}\ \bibnamefont {Ehrlich}}, \bibinfo {author}
  {\bibfnamefont {L.~L.}\ \bibnamefont {Erskine}}, \bibinfo {author}
  {\bibfnamefont {A.~A.}\ \bibnamefont {Heikal}}, \bibinfo {author}
  {\bibfnamefont {S.~M.}\ \bibnamefont {Kuebler}}, \bibinfo {author}
  {\bibfnamefont {I.-Y.~S.}\ \bibnamefont {Lee}}, \bibinfo {author}
  {\bibfnamefont {D.}~\bibnamefont {McCord-Maughon}}, \bibinfo {author}
  {\bibfnamefont {J.}~\bibnamefont {Qin}}, \bibinfo {author} {\bibfnamefont
  {H.}~\bibnamefont {Rockel}}, \bibinfo {author} {\bibfnamefont
  {M.}~\bibnamefont {Rumi}}, \bibinfo {author} {\bibfnamefont {X.-L.}\
  \bibnamefont {Wu}}, \bibinfo {author} {\bibfnamefont {S.}~\bibnamefont
  {Marder}}, \ and\ \bibinfo {author} {\bibfnamefont {J.~W.}\ \bibnamefont
  {Perry}},\ }\href@noop {} {\bibfield  {journal} {\bibinfo  {journal}
  {Nature},\ }\textbf {\bibinfo {volume} {398}},\ \bibinfo {pages} {51}
  (\bibinfo {year} {1999})}\BibitemShut {NoStop}%
\bibitem [{\citenamefont {Kawata}\ \emph {et~al.}(2001)\citenamefont {Kawata},
  \citenamefont {Sun}, \citenamefont {Tanaka},\ and\ \citenamefont
  {Takada}}]{kawat01.01}%
  \BibitemOpen
  \bibfield  {author} {\bibinfo {author} {\bibfnamefont {S.}~\bibnamefont
  {Kawata}}, \bibinfo {author} {\bibfnamefont {H.-B.}\ \bibnamefont {Sun}},
  \bibinfo {author} {\bibfnamefont {T.}~\bibnamefont {Tanaka}}, \ and\ \bibinfo
  {author} {\bibfnamefont {K.}~\bibnamefont {Takada}},\ }\href@noop {}
  {\bibfield  {journal} {\bibinfo  {journal} {Nature},\ }\textbf {\bibinfo
  {volume} {412}},\ \bibinfo {pages} {697} (\bibinfo {year}
  {2001})}\BibitemShut {NoStop}%
\bibitem [{\citenamefont {Karotki}\ \emph {et~al.}(2004)\citenamefont
  {Karotki}, \citenamefont {Drobizhev}, \citenamefont {Dzenis}, \citenamefont
  {Taylor}, \citenamefont {Anderson},\ and\ \citenamefont
  {Rebane}}]{karot04.01}%
  \BibitemOpen
  \bibfield  {author} {\bibinfo {author} {\bibfnamefont {A.}~\bibnamefont
  {Karotki}}, \bibinfo {author} {\bibfnamefont {M.}~\bibnamefont {Drobizhev}},
  \bibinfo {author} {\bibfnamefont {Y.}~\bibnamefont {Dzenis}}, \bibinfo
  {author} {\bibfnamefont {P.~N.}\ \bibnamefont {Taylor}}, \bibinfo {author}
  {\bibfnamefont {H.~L.}\ \bibnamefont {Anderson}}, \ and\ \bibinfo {author}
  {\bibfnamefont {A.}~\bibnamefont {Rebane}},\ }\href@noop {} {\bibfield
  {journal} {\bibinfo  {journal} {Phys. Chem. Chem. Phys.},\ }\textbf {\bibinfo
  {volume} {6}},\ \bibinfo {pages} {7} (\bibinfo {year} {2004})}\BibitemShut
  {NoStop}%
\bibitem [{\citenamefont {Roy}\ \emph {et~al.}(2003)\citenamefont {Roy},
  \citenamefont {Ohulchanskyy}, \citenamefont {Pudavar}, \citenamefont
  {Bergey}, \citenamefont {Oseroff}, \citenamefont {Morgan}, \citenamefont
  {Dougherty},\ and\ \citenamefont {Prasad}}]{roy03.01}%
  \BibitemOpen
  \bibfield  {author} {\bibinfo {author} {\bibfnamefont {I.}~\bibnamefont
  {Roy}}, \bibinfo {author} {\bibfnamefont {T.~Y.}\ \bibnamefont
  {Ohulchanskyy}}, \bibinfo {author} {\bibfnamefont {H.~E.}\ \bibnamefont
  {Pudavar}}, \bibinfo {author} {\bibfnamefont {E.~J.}\ \bibnamefont {Bergey}},
  \bibinfo {author} {\bibfnamefont {A.~R.}\ \bibnamefont {Oseroff}}, \bibinfo
  {author} {\bibfnamefont {J.}~\bibnamefont {Morgan}}, \bibinfo {author}
  {\bibfnamefont {T.~J.}\ \bibnamefont {Dougherty}}, \ and\ \bibinfo {author}
  {\bibfnamefont {P.~N.}\ \bibnamefont {Prasad}},\ }\href@noop {} {\bibfield
  {journal} {\bibinfo  {journal} {J. Am. Chem. Soc.},\ }\textbf {\bibinfo
  {volume} {125}},\ \bibinfo {pages} {7860} (\bibinfo {year}
  {2003})}\BibitemShut {NoStop}%
\bibitem [{\citenamefont {Bredas}\ \emph {et~al.}(1994)\citenamefont {Bredas},
  \citenamefont {Adant}, \citenamefont {Persoons},\ and\ \citenamefont
  {Pierc}}]{Breda94.01}%
  \BibitemOpen
  \bibfield  {author} {\bibinfo {author} {\bibfnamefont {J.~L.}\ \bibnamefont
  {Bredas}}, \bibinfo {author} {\bibfnamefont {C.}~\bibnamefont {Adant}},
  \bibinfo {author} {\bibfnamefont {A.}~\bibnamefont {Persoons}}, \ and\
  \bibinfo {author} {\bibnamefont {Pierc}},\ }\href@noop {} {\bibfield
  {journal} {\bibinfo  {journal} {Chem. R},\ }\textbf {\bibinfo {volume}
  {94}},\ \bibinfo {pages} {243} (\bibinfo {year} {1994})}\BibitemShut
  {NoStop}%
\bibitem [{\citenamefont {Albota}\ \emph {et~al.}(1998)\citenamefont {Albota},
  \citenamefont {Beljonne}, \citenamefont {Bredas}, \citenamefont {Ehrlich},
  \citenamefont {Fu}, \citenamefont {Heikal}, \citenamefont {Hess},
  \citenamefont {Kogej}, \citenamefont {Levin}, \citenamefont {Marder},
  \citenamefont {McCord-Maughon}, \citenamefont {Perry}, \citenamefont
  {Rockel}, \citenamefont {Rumi}, \citenamefont {Subramaniam},\ and\
  \citenamefont {Webb}}]{Albot98.01}%
  \BibitemOpen
  \bibfield  {author} {\bibinfo {author} {\bibfnamefont {M.}~\bibnamefont
  {Albota}}, \bibinfo {author} {\bibfnamefont {D.}~\bibnamefont {Beljonne}},
  \bibinfo {author} {\bibfnamefont {J.~L.}\ \bibnamefont {Bredas}}, \bibinfo
  {author} {\bibfnamefont {J.~E.}\ \bibnamefont {Ehrlich}}, \bibinfo {author}
  {\bibfnamefont {J.-Y.}\ \bibnamefont {Fu}}, \bibinfo {author} {\bibfnamefont
  {A.~A.}\ \bibnamefont {Heikal}}, \bibinfo {author} {\bibfnamefont {S.~E.}\
  \bibnamefont {Hess}}, \bibinfo {author} {\bibfnamefont {T.}~\bibnamefont
  {Kogej}}, \bibinfo {author} {\bibfnamefont {M.~D.}\ \bibnamefont {Levin}},
  \bibinfo {author} {\bibfnamefont {S.~R.}\ \bibnamefont {Marder}}, \bibinfo
  {author} {\bibfnamefont {D.}~\bibnamefont {McCord-Maughon}}, \bibinfo
  {author} {\bibfnamefont {J.~W.}\ \bibnamefont {Perry}}, \bibinfo {author}
  {\bibfnamefont {H.}~\bibnamefont {Rockel}}, \bibinfo {author} {\bibfnamefont
  {M.}~\bibnamefont {Rumi}}, \bibinfo {author} {\bibfnamefont {G.}~\bibnamefont
  {Subramaniam}}, \ and\ \bibinfo {author} {\bibfnamefont {W.~W.}\ \bibnamefont
  {Webb}},\ }\href@noop {} {\bibfield  {journal} {\bibinfo  {journal}
  {Science},\ }\textbf {\bibinfo {volume} {281}},\ \bibinfo {pages} {1653}
  (\bibinfo {year} {1998})}\BibitemShut {NoStop}%
\bibitem [{\citenamefont {Fu}\ \emph {et~al.}(2007){\natexlab{a}}\citenamefont
  {Fu}, \citenamefont {Lazaro}, \citenamefont {Hagan}, \citenamefont
  {Van~Stryland}, \citenamefont {Przhonska}, \citenamefont {Bondar},
  \citenamefont {Slominsky},\ and\ \citenamefont {Kachkovski}}]{fu07.01}%
  \BibitemOpen
  \bibfield  {author} {\bibinfo {author} {\bibfnamefont {J.}~\bibnamefont
  {Fu}}, \bibinfo {author} {\bibfnamefont {A.~P.}\ \bibnamefont {Lazaro}},
  \bibinfo {author} {\bibfnamefont {D.~J.}\ \bibnamefont {Hagan}}, \bibinfo
  {author} {\bibfnamefont {E.~W.}\ \bibnamefont {Van~Stryland}}, \bibinfo
  {author} {\bibfnamefont {O.~V.}\ \bibnamefont {Przhonska}}, \bibinfo {author}
  {\bibfnamefont {M.~V.}\ \bibnamefont {Bondar}}, \bibinfo {author}
  {\bibfnamefont {Y.~L.}\ \bibnamefont {Slominsky}}, \ and\ \bibinfo {author}
  {\bibfnamefont {A.~D.}\ \bibnamefont {Kachkovski}},\ }\href@noop {}
  {\bibfield  {journal} {\bibinfo  {journal} {J. Opt. Soc. Am. B},\ }\textbf
  {\bibinfo {volume} {24}},\ \bibinfo {pages} {56} (\bibinfo {year}
  {2007}{\natexlab{a}})}\BibitemShut {NoStop}%
\bibitem [{\citenamefont {Rumi}\ \emph {et~al.}(2000)\citenamefont {Rumi},
  \citenamefont {Ehrlich}, \citenamefont {Heikal}, \citenamefont {Perry},
  \citenamefont {Barlow}, \citenamefont {Hu}, \citenamefont {McCord-Maughon},
  \citenamefont {Parker}, \citenamefont {Rockel}, \citenamefont {Thayumanavan},
  \citenamefont {Marder}, \citenamefont {Beljonne},\ and\ \citenamefont
  {Bredas}}]{Rumi00.01}%
  \BibitemOpen
  \bibfield  {author} {\bibinfo {author} {\bibfnamefont {M.}~\bibnamefont
  {Rumi}}, \bibinfo {author} {\bibfnamefont {J.~E.}\ \bibnamefont {Ehrlich}},
  \bibinfo {author} {\bibfnamefont {A.~A.}\ \bibnamefont {Heikal}}, \bibinfo
  {author} {\bibfnamefont {J.~W.}\ \bibnamefont {Perry}}, \bibinfo {author}
  {\bibfnamefont {S.}~\bibnamefont {Barlow}}, \bibinfo {author} {\bibfnamefont
  {Z.}~\bibnamefont {Hu}}, \bibinfo {author} {\bibfnamefont {D.}~\bibnamefont
  {McCord-Maughon}}, \bibinfo {author} {\bibfnamefont {T.~C.}\ \bibnamefont
  {Parker}}, \bibinfo {author} {\bibfnamefont {H.}~\bibnamefont {Rockel}},
  \bibinfo {author} {\bibfnamefont {S.}~\bibnamefont {Thayumanavan}}, \bibinfo
  {author} {\bibfnamefont {S.~R.}\ \bibnamefont {Marder}}, \bibinfo {author}
  {\bibfnamefont {D.}~\bibnamefont {Beljonne}}, \ and\ \bibinfo {author}
  {\bibfnamefont {J.~L.}\ \bibnamefont {Bredas}},\ }\href@noop {} {\bibfield
  {journal} {\bibinfo  {journal} {J. Am. Chem. Soc.},\ }\textbf {\bibinfo
  {volume} {122}},\ \bibinfo {pages} {9500} (\bibinfo {year}
  {2000})}\BibitemShut {NoStop}%
\bibitem [{\citenamefont {Liao}\ \emph {et~al.}(2005)\citenamefont {Liao},
  \citenamefont {Eichinger}, \citenamefont {Firestone}, \citenamefont {Haller},
  \citenamefont {Luo}, \citenamefont {Kaminsky}, \citenamefont {Benedict},
  \citenamefont {Reid},\ and\ \citenamefont {Jen}}]{liao05.01}%
  \BibitemOpen
  \bibfield  {author} {\bibinfo {author} {\bibfnamefont {Y.}~\bibnamefont
  {Liao}}, \bibinfo {author} {\bibfnamefont {B.~E.}\ \bibnamefont {Eichinger}},
  \bibinfo {author} {\bibfnamefont {K.~A.}\ \bibnamefont {Firestone}}, \bibinfo
  {author} {\bibfnamefont {M.}~\bibnamefont {Haller}}, \bibinfo {author}
  {\bibfnamefont {J.}~\bibnamefont {Luo}}, \bibinfo {author} {\bibfnamefont
  {W.}~\bibnamefont {Kaminsky}}, \bibinfo {author} {\bibfnamefont {J.~B.}\
  \bibnamefont {Benedict}}, \bibinfo {author} {\bibfnamefont {P.~J.}\
  \bibnamefont {Reid}}, \ and\ \bibinfo {author} {\bibfnamefont {A.~K.~Y.}\
  \bibnamefont {Jen}},\ }\href@noop {} {\bibfield  {journal} {\bibinfo
  {journal} {J. Am. Chem. Soc.},\ }\textbf {\bibinfo {volume} {127}},\ \bibinfo
  {pages} {2758} (\bibinfo {year} {2005})}\BibitemShut {NoStop}%
\bibitem [{\citenamefont {Nandi}\ \emph {et~al.}(2008)\citenamefont {Nandi},
  \citenamefont {Panja},\ and\ \citenamefont {Ghanty}}]{nandi08.01}%
  \BibitemOpen
  \bibfield  {author} {\bibinfo {author} {\bibfnamefont {P.~K.}\ \bibnamefont
  {Nandi}}, \bibinfo {author} {\bibfnamefont {N.}~\bibnamefont {Panja}}, \ and\
  \bibinfo {author} {\bibfnamefont {T.~K.}\ \bibnamefont {Ghanty}},\
  }\href@noop {} {\bibfield  {journal} {\bibinfo  {journal} {J. Phys. Chem.
  A},\ }\textbf {\bibinfo {volume} {112}},\ \bibinfo {pages} {4844} (\bibinfo
  {year} {2008})}\BibitemShut {NoStop}%
\bibitem [{\citenamefont {Kuzyk}(2009)}]{kuzyk09.01}%
  \BibitemOpen
  \bibfield  {author} {\bibinfo {author} {\bibfnamefont {M.~G.}\ \bibnamefont
  {Kuzyk}},\ }\href@noop {} {\bibfield  {journal} {\bibinfo  {journal} {J. Mat.
  Chem.},\ }\textbf {\bibinfo {volume} {19}},\ \bibinfo {pages} {7444}
  (\bibinfo {year} {2009})}\BibitemShut {NoStop}%
\bibitem [{\citenamefont {Kuzyk}(2010)}]{kuzyk10.01}%
  \BibitemOpen
  \bibfield  {author} {\bibinfo {author} {\bibfnamefont {M.~G.}\ \bibnamefont
  {Kuzyk}},\ }\href@noop {} {\bibfield  {journal} {\bibinfo  {journal}
  {Nonliner Optics Quantum Optics},\ }\textbf {\bibinfo {volume} {40}},\
  \bibinfo {pages} {1} (\bibinfo {year} {2010})}\BibitemShut {NoStop}%
\bibitem [{\citenamefont {Chernyak}\ \emph {et~al.}(2000)\citenamefont
  {Chernyak}, \citenamefont {Tretiak},\ and\ \citenamefont
  {Mukamel}}]{chern20.01}%
  \BibitemOpen
  \bibfield  {author} {\bibinfo {author} {\bibfnamefont {V.}~\bibnamefont
  {Chernyak}}, \bibinfo {author} {\bibfnamefont {S.}~\bibnamefont {Tretiak}}, \
  and\ \bibinfo {author} {\bibnamefont {Mukamel}},\ }\href@noop {} {\bibfield
  {journal} {\bibinfo  {journal} {Chem. Phys. Lett.},\ }\textbf {\bibinfo
  {volume} {319}},\ \bibinfo {pages} {261} (\bibinfo {year}
  {2000})}\BibitemShut {NoStop}%
\bibitem [{\citenamefont {Tripathy}\ \emph {et~al.}(2004)\citenamefont
  {Tripathy}, \citenamefont {P\'{e}rez~Moreno}, \citenamefont {Kuzyk},
  \citenamefont {Coe}, \citenamefont {Clays},\ and\ \citenamefont
  {Kelley}}]{Tripa04.01}%
  \BibitemOpen
  \bibfield  {author} {\bibinfo {author} {\bibfnamefont {K.}~\bibnamefont
  {Tripathy}}, \bibinfo {author} {\bibfnamefont {J.}~\bibnamefont
  {P\'{e}rez~Moreno}}, \bibinfo {author} {\bibfnamefont {M.~G.}\ \bibnamefont
  {Kuzyk}}, \bibinfo {author} {\bibfnamefont {B.~J.}\ \bibnamefont {Coe}},
  \bibinfo {author} {\bibfnamefont {K.}~\bibnamefont {Clays}}, \ and\ \bibinfo
  {author} {\bibfnamefont {A.~M.}\ \bibnamefont {Kelley}},\ }\href@noop {}
  {\bibfield  {journal} {\bibinfo  {journal} {J. Chem. Phys.},\ }\textbf
  {\bibinfo {volume} {121}},\ \bibinfo {pages} {7932} (\bibinfo {year}
  {2004})}\BibitemShut {NoStop}%
\bibitem [{\citenamefont {Tripathy}\ \emph {et~al.}(2006)\citenamefont
  {Tripathy}, \citenamefont {P\'{e}rez~Moreno}, \citenamefont {Kuzyk},
  \citenamefont {Coe}, \citenamefont {Clays},\ and\ \citenamefont
  {Kelley}}]{tripa07.01}%
  \BibitemOpen
  \bibfield  {author} {\bibinfo {author} {\bibfnamefont {K.}~\bibnamefont
  {Tripathy}}, \bibinfo {author} {\bibfnamefont {J.}~\bibnamefont
  {P\'{e}rez~Moreno}}, \bibinfo {author} {\bibfnamefont {M.~G.}\ \bibnamefont
  {Kuzyk}}, \bibinfo {author} {\bibfnamefont {B.~J.}\ \bibnamefont {Coe}},
  \bibinfo {author} {\bibfnamefont {K.}~\bibnamefont {Clays}}, \ and\ \bibinfo
  {author} {\bibfnamefont {A.~M.}\ \bibnamefont {Kelley}},\ }\href@noop {}
  {\bibfield  {journal} {\bibinfo  {journal} {J . Chem. Phys.},\ }\textbf
  {\bibinfo {volume} {125}},\ \bibinfo {pages} {079905} (\bibinfo {year}
  {2006})}\BibitemShut {NoStop}%
\bibitem [{\citenamefont {Bishop}\ \emph {et~al.}(2000)\citenamefont {Bishop},
  \citenamefont {Champagne},\ and\ \citenamefont {Kirtman}}]{bisho20.01}%
  \BibitemOpen
  \bibfield  {author} {\bibinfo {author} {\bibfnamefont {D.~M.}\ \bibnamefont
  {Bishop}}, \bibinfo {author} {\bibfnamefont {B.}~\bibnamefont {Champagne}}, \
  and\ \bibinfo {author} {\bibfnamefont {B.}~\bibnamefont {Kirtman}},\
  }\href@noop {} {\bibfield  {journal} {\bibinfo  {journal} {Chem. Phys.
  Lett.},\ }\textbf {\bibinfo {volume} {329}},\ \bibinfo {pages} {329}
  (\bibinfo {year} {2000})}\BibitemShut {NoStop}%
\bibitem [{\citenamefont {Das}\ \emph {et~al.}(1993)\citenamefont {Das},
  \citenamefont {Yeates},\ and\ \citenamefont {Dudis}}]{das93.01}%
  \BibitemOpen
  \bibfield  {author} {\bibinfo {author} {\bibfnamefont {G.~P.}\ \bibnamefont
  {Das}}, \bibinfo {author} {\bibfnamefont {A.~T.}\ \bibnamefont {Yeates}}, \
  and\ \bibinfo {author} {\bibfnamefont {D.}~\bibnamefont {Dudis}},\
  }\href@noop {} {\bibfield  {journal} {\bibinfo  {journal} {Chem. Phys.
  Lett.},\ }\textbf {\bibinfo {volume} {212}},\ \bibinfo {pages} {671}
  (\bibinfo {year} {1993})}\BibitemShut {NoStop}%
\bibitem [{\citenamefont {Kirtman}\ and\ \citenamefont
  {Champagne}(1997)}]{kirtm97.01}%
  \BibitemOpen
  \bibfield  {author} {\bibinfo {author} {\bibfnamefont {B.}~\bibnamefont
  {Kirtman}}\ and\ \bibinfo {author} {\bibfnamefont {B.}~\bibnamefont
  {Champagne}},\ }\href@noop {} {\bibfield  {journal} {\bibinfo  {journal}
  {Int. Rev. Phys. Chem.},\ }\textbf {\bibinfo {volume} {16}},\ \bibinfo
  {pages} {389} (\bibinfo {year} {1997})}\BibitemShut {NoStop}%
\bibitem [{\citenamefont {Painelli}(1998)}]{paine98.01}%
  \BibitemOpen
  \bibfield  {author} {\bibinfo {author} {\bibfnamefont {A.}~\bibnamefont
  {Painelli}},\ }\href@noop {} {\bibfield  {journal} {\bibinfo  {journal}
  {Chem. Phys. Lett.},\ }\textbf {\bibinfo {volume} {285}},\ \bibinfo {pages}
  {352} (\bibinfo {year} {1998})}\BibitemShut {NoStop}%
\bibitem [{\citenamefont {Bishop}\ \emph {et~al.}(1998)\citenamefont {Bishop},
  \citenamefont {Champagne},\ and\ \citenamefont {Kirtman}}]{bisho98.01}%
  \BibitemOpen
  \bibfield  {author} {\bibinfo {author} {\bibfnamefont {D.~M.}\ \bibnamefont
  {Bishop}}, \bibinfo {author} {\bibfnamefont {B.}~\bibnamefont {Champagne}}, \
  and\ \bibinfo {author} {\bibfnamefont {B.}~\bibnamefont {Kirtman}},\
  }\href@noop {} {\bibfield  {journal} {\bibinfo  {journal} {J. Chem. Phys.},\
  }\textbf {\bibinfo {volume} {109}},\ \bibinfo {pages} {9987} (\bibinfo {year}
  {1998})}\BibitemShut {NoStop}%
\bibitem [{\citenamefont {Clays}\ and\ \citenamefont
  {Persoons}(1991)}]{clays91.01}%
  \BibitemOpen
  \bibfield  {author} {\bibinfo {author} {\bibfnamefont {K.}~\bibnamefont
  {Clays}}\ and\ \bibinfo {author} {\bibfnamefont {A.}~\bibnamefont
  {Persoons}},\ }\href@noop {} {\bibfield  {journal} {\bibinfo  {journal}
  {Phys. Rev. Lett.},\ }\textbf {\bibinfo {volume} {66}},\ \bibinfo {pages}
  {2980} (\bibinfo {year} {1991})}\BibitemShut {NoStop}%
\bibitem [{\citenamefont {Oudar}\ and\ \citenamefont
  {Chemla}(1977)}]{oudar77.03}%
  \BibitemOpen
  \bibfield  {author} {\bibinfo {author} {\bibfnamefont {J.~L.}\ \bibnamefont
  {Oudar}}\ and\ \bibinfo {author} {\bibfnamefont {D.~S.}\ \bibnamefont
  {Chemla}},\ }\href@noop {} {\bibfield  {journal} {\bibinfo  {journal} {J.
  Chem Phys.},\ }\textbf {\bibinfo {volume} {66}},\ \bibinfo {pages} {2664}
  (\bibinfo {year} {1977})}\BibitemShut {NoStop}%
\bibitem [{\citenamefont {Orr}\ and\ \citenamefont {Ward}(1971)}]{orr71.01}%
  \BibitemOpen
  \bibfield  {author} {\bibinfo {author} {\bibfnamefont {B.~J.}\ \bibnamefont
  {Orr}}\ and\ \bibinfo {author} {\bibfnamefont {J.~F.}\ \bibnamefont {Ward}},\
  }\href@noop {} {\bibfield  {journal} {\bibinfo  {journal} {Molecular
  Physics},\ }\textbf {\bibinfo {volume} {20}},\ \bibinfo {pages} {513}
  (\bibinfo {year} {1971})}\BibitemShut {NoStop}%
\bibitem [{\citenamefont {Bethe}\ and\ \citenamefont
  {Salpeter}(1977)}]{Bethe77.01}%
  \BibitemOpen
  \bibfield  {author} {\bibinfo {author} {\bibfnamefont {H.~A.}\ \bibnamefont
  {Bethe}}\ and\ \bibinfo {author} {\bibfnamefont {E.~E.}\ \bibnamefont
  {Salpeter}},\ }\href@noop {} {\emph {\bibinfo {title} {Quantum Mechanics of
  One and Two Electron Atoms}}}\ (\bibinfo  {publisher} {Plenum},\ \bibinfo
  {address} {New York},\ \bibinfo {year} {1977})\BibitemShut {NoStop}%
\bibitem [{\citenamefont {Kuzyk}(2000){\natexlab{a}}}]{kuzyk00.01}%
  \BibitemOpen
  \bibfield  {author} {\bibinfo {author} {\bibfnamefont {M.~G.}\ \bibnamefont
  {Kuzyk}},\ }\href@noop {} {\bibfield  {journal} {\bibinfo  {journal} {Phys.
  Rev. Lett.},\ }\textbf {\bibinfo {volume} {85}},\ \bibinfo {pages} {1218}
  (\bibinfo {year} {2000}{\natexlab{a}})}\BibitemShut {NoStop}%
\bibitem [{\citenamefont {Kuzyk}(2000){\natexlab{b}}}]{kuzyk00.02}%
  \BibitemOpen
  \bibfield  {author} {\bibinfo {author} {\bibfnamefont {M.~G.}\ \bibnamefont
  {Kuzyk}},\ }\href@noop {} {\bibfield  {journal} {\bibinfo  {journal} {Opt.
  Lett.},\ }\textbf {\bibinfo {volume} {25}},\ \bibinfo {pages} {1183}
  (\bibinfo {year} {2000}{\natexlab{b}})}\BibitemShut {NoStop}%
\bibitem [{\citenamefont {Kuzyk}(2001)}]{kuzyk01.01}%
  \BibitemOpen
  \bibfield  {author} {\bibinfo {author} {\bibfnamefont {M.~G.}\ \bibnamefont
  {Kuzyk}},\ }\href@noop {} {\bibfield  {journal} {\bibinfo  {journal} {IEEE
  Journal on Selected Topics in Quantum Electronics},\ }\textbf {\bibinfo
  {volume} {7}},\ \bibinfo {pages} {774 } (\bibinfo {year} {2001})}\BibitemShut
  {NoStop}%
\bibitem [{\citenamefont {Kuzyk}(2003){\natexlab{a}}}]{kuzyk03.01}%
  \BibitemOpen
  \bibfield  {author} {\bibinfo {author} {\bibfnamefont {M.~G.}\ \bibnamefont
  {Kuzyk}},\ }\href@noop {} {\bibfield  {journal} {\bibinfo  {journal} {Opt.
  Lett.},\ }\textbf {\bibinfo {volume} {28}},\ \bibinfo {pages} {135} (\bibinfo
  {year} {2003}{\natexlab{a}})}\BibitemShut {NoStop}%
\bibitem [{\citenamefont {Kuzyk}(2003){\natexlab{b}}}]{kuzyk03.02}%
  \BibitemOpen
  \bibfield  {author} {\bibinfo {author} {\bibfnamefont {M.~G.}\ \bibnamefont
  {Kuzyk}},\ }\href@noop {} {\bibfield  {journal} {\bibinfo  {journal} {Phys.
  Rev. Lett.},\ }\textbf {\bibinfo {volume} {90}},\ \bibinfo {pages} {039902}
  (\bibinfo {year} {2003}{\natexlab{b}})}\BibitemShut {NoStop}%
\bibitem [{\citenamefont {Kuzyk}(2004)}]{kuzyk04.02}%
  \BibitemOpen
  \bibfield  {author} {\bibinfo {author} {\bibfnamefont {M.~G.}\ \bibnamefont
  {Kuzyk}},\ }\href@noop {} {\bibfield  {journal} {\bibinfo  {journal} {J.
  Nonl. Opt. Phys. \& Mat.},\ }\textbf {\bibinfo {volume} {13}},\ \bibinfo
  {pages} {461} (\bibinfo {year} {2004})}\BibitemShut {NoStop}%
\bibitem [{\citenamefont {Zhou}\ \emph {et~al.}(2006)\citenamefont {Zhou},
  \citenamefont {Kuzyk},\ and\ \citenamefont {Watkins}}]{zhou06.01}%
  \BibitemOpen
  \bibfield  {author} {\bibinfo {author} {\bibfnamefont {J.}~\bibnamefont
  {Zhou}}, \bibinfo {author} {\bibfnamefont {M.~G.}\ \bibnamefont {Kuzyk}}, \
  and\ \bibinfo {author} {\bibfnamefont {D.~S.}\ \bibnamefont {Watkins}},\
  }\href@noop {} {\bibfield  {journal} {\bibinfo  {journal} {Opt. Lett.},\
  }\textbf {\bibinfo {volume} {31}},\ \bibinfo {pages} {2891} (\bibinfo {year}
  {2006})}\BibitemShut {NoStop}%
\bibitem [{\citenamefont {Zhou}\ \emph
  {et~al.}(2007){\natexlab{a}}\citenamefont {Zhou}, \citenamefont {Szafruga},
  \citenamefont {Watkins},\ and\ \citenamefont {Kuzyk}}]{zhou07.02}%
  \BibitemOpen
  \bibfield  {author} {\bibinfo {author} {\bibfnamefont {J.}~\bibnamefont
  {Zhou}}, \bibinfo {author} {\bibfnamefont {U.~B.}\ \bibnamefont {Szafruga}},
  \bibinfo {author} {\bibfnamefont {D.~S.}\ \bibnamefont {Watkins}}, \ and\
  \bibinfo {author} {\bibfnamefont {M.~G.}\ \bibnamefont {Kuzyk}},\ }\href@noop
  {} {\bibfield  {journal} {\bibinfo  {journal} {Phys. Rev. A},\ }\textbf
  {\bibinfo {volume} {76}},\ \bibinfo {pages} {053831} (\bibinfo {year}
  {2007}{\natexlab{a}})}\BibitemShut {NoStop}%
\bibitem [{\citenamefont {P\'{e}rez-Moreno}\ \emph {et~al.}(2007)\citenamefont
  {P\'{e}rez-Moreno}, \citenamefont {Zhao}, \citenamefont {Clays},\ and\
  \citenamefont {Kuzyk}}]{perez07.01}%
  \BibitemOpen
  \bibfield  {author} {\bibinfo {author} {\bibfnamefont {J.}~\bibnamefont
  {P\'{e}rez-Moreno}}, \bibinfo {author} {\bibfnamefont {Y.}~\bibnamefont
  {Zhao}}, \bibinfo {author} {\bibfnamefont {K.}~\bibnamefont {Clays}}, \ and\
  \bibinfo {author} {\bibfnamefont {M.~G.}\ \bibnamefont {Kuzyk}},\ }\href@noop
  {} {\bibfield  {journal} {\bibinfo  {journal} {Opt. Lett.},\ }\textbf
  {\bibinfo {volume} {32}},\ \bibinfo {pages} {59} (\bibinfo {year}
  {2007})}\BibitemShut {NoStop}%
\bibitem [{\citenamefont {P\'{e}rez-Moreno}\ \emph {et~al.}(2009)\citenamefont
  {P\'{e}rez-Moreno}, \citenamefont {Zhao}, \citenamefont {Clays},
  \citenamefont {Kuzyk}, \citenamefont {Shen}, \citenamefont {Qiu},
  \citenamefont {Hao},\ and\ \citenamefont {Guo}}]{perez09.01}%
  \BibitemOpen
  \bibfield  {author} {\bibinfo {author} {\bibfnamefont {J.}~\bibnamefont
  {P\'{e}rez-Moreno}}, \bibinfo {author} {\bibfnamefont {Y.}~\bibnamefont
  {Zhao}}, \bibinfo {author} {\bibfnamefont {K.}~\bibnamefont {Clays}},
  \bibinfo {author} {\bibfnamefont {M.~G.}\ \bibnamefont {Kuzyk}}, \bibinfo
  {author} {\bibfnamefont {Y.}~\bibnamefont {Shen}}, \bibinfo {author}
  {\bibfnamefont {L.}~\bibnamefont {Qiu}}, \bibinfo {author} {\bibfnamefont
  {J.}~\bibnamefont {Hao}}, \ and\ \bibinfo {author} {\bibfnamefont
  {K.}~\bibnamefont {Guo}},\ }\href@noop {} {\bibfield  {journal} {\bibinfo
  {journal} {J. Am. Chem. Soc.},\ }\textbf {\bibinfo {volume} {131}},\ \bibinfo
  {pages} {5084–5093} (\bibinfo {year} {2009})}\BibitemShut {NoStop}%
\bibitem [{\citenamefont {Kuzyk}(2006){\natexlab{a}}}]{kuzyk06.01}%
  \BibitemOpen
  \bibfield  {author} {\bibinfo {author} {\bibfnamefont {M.~G.}\ \bibnamefont
  {Kuzyk}},\ }\href@noop {} {\bibfield  {journal} {\bibinfo  {journal} {J.
  Nonl. Opt. Phys. \& Mat.},\ }\textbf {\bibinfo {volume} {15}},\ \bibinfo
  {pages} {77} (\bibinfo {year} {2006}{\natexlab{a}})}\BibitemShut {NoStop}%
\bibitem [{\citenamefont {Watkins}\ and\ \citenamefont
  {Kuzyk}(2009)}]{watkins09.01}%
  \BibitemOpen
  \bibfield  {author} {\bibinfo {author} {\bibfnamefont {D.~S.}\ \bibnamefont
  {Watkins}}\ and\ \bibinfo {author} {\bibfnamefont {M.~G.}\ \bibnamefont
  {Kuzyk}},\ }\href@noop {} {\bibfield  {journal} {\bibinfo  {journal} {J.
  Chem. Phys.},\ }\textbf {\bibinfo {volume} {131}},\ \bibinfo {pages} {064110}
  (\bibinfo {year} {2009})}\BibitemShut {NoStop}%
\bibitem [{\citenamefont {Kuzyk}\ and\ \citenamefont
  {Kuzyk}(2008)}]{kuzyk08.01}%
  \BibitemOpen
  \bibfield  {author} {\bibinfo {author} {\bibfnamefont {M.~C.}\ \bibnamefont
  {Kuzyk}}\ and\ \bibinfo {author} {\bibfnamefont {M.~G.}\ \bibnamefont
  {Kuzyk}},\ }\href@noop {} {\bibfield  {journal} {\bibinfo  {journal} {J. Opt.
  Soc. Am. B.},\ }\textbf {\bibinfo {volume} {25}},\ \bibinfo {pages} {103}
  (\bibinfo {year} {2008})}\BibitemShut {NoStop}%
\bibitem [{\citenamefont {Champagne}\ and\ \citenamefont
  {Kirtman}(2005)}]{Champ05.01}%
  \BibitemOpen
  \bibfield  {author} {\bibinfo {author} {\bibfnamefont {B.}~\bibnamefont
  {Champagne}}\ and\ \bibinfo {author} {\bibfnamefont {B.}~\bibnamefont
  {Kirtman}},\ }\href@noop {} {\bibfield  {journal} {\bibinfo  {journal} {Phys.
  Rev. Lett.},\ }\textbf {\bibinfo {volume} {95}},\ \bibinfo {pages} {109401}
  (\bibinfo {year} {2005})}\BibitemShut {NoStop}%
\bibitem [{\citenamefont {Lagarias}\ \emph {et~al.}(1998)\citenamefont
  {Lagarias}, \citenamefont {Reeds}, \citenamefont {Wright},\ and\
  \citenamefont {Wright}}]{lagar98.01}%
  \BibitemOpen
  \bibfield  {author} {\bibinfo {author} {\bibfnamefont {J.~C.}\ \bibnamefont
  {Lagarias}}, \bibinfo {author} {\bibfnamefont {J.~A.}\ \bibnamefont {Reeds}},
  \bibinfo {author} {\bibfnamefont {M.~H.}\ \bibnamefont {Wright}}, \ and\
  \bibinfo {author} {\bibfnamefont {P.}~\bibnamefont {Wright}},\ }\href@noop {}
  {\bibfield  {journal} {\bibinfo  {journal} {SIAM J. Optim.},\ }\textbf
  {\bibinfo {volume} {9}},\ \bibinfo {pages} {112} (\bibinfo {year}
  {1998})}\BibitemShut {NoStop}%
\bibitem [{\citenamefont {Kuzyk}(2005)}]{kuzyk05.02}%
  \BibitemOpen
  \bibfield  {author} {\bibinfo {author} {\bibfnamefont {M.~G.}\ \bibnamefont
  {Kuzyk}},\ }\href@noop {} {\bibfield  {journal} {\bibinfo  {journal} {Phys.
  Rev. A},\ }\textbf {\bibinfo {volume} {72}},\ \bibinfo {pages} {053819}
  (\bibinfo {year} {2005})}\BibitemShut {NoStop}%
\bibitem [{\citenamefont {Zhou}\ \emph
  {et~al.}(2007){\natexlab{b}}\citenamefont {Zhou}, \citenamefont {Kuzyk},\
  and\ \citenamefont {Watkins}}]{zhou07.01}%
  \BibitemOpen
  \bibfield  {author} {\bibinfo {author} {\bibfnamefont {J.}~\bibnamefont
  {Zhou}}, \bibinfo {author} {\bibfnamefont {M.~G.}\ \bibnamefont {Kuzyk}}, \
  and\ \bibinfo {author} {\bibfnamefont {D.~S.}\ \bibnamefont {Watkins}},\
  }\href@noop {} {\bibfield  {journal} {\bibinfo  {journal} {Opt. Lett.},\
  }\textbf {\bibinfo {volume} {32}},\ \bibinfo {pages} {944} (\bibinfo {year}
  {2007}{\natexlab{b}})}\BibitemShut {NoStop}%
\bibitem [{\citenamefont {Zienkiewicz}\ \emph {et~al.}(2005)\citenamefont
  {Zienkiewicz}, \citenamefont {Taylor},\ and\ \citenamefont
  {Zhu}}]{zienk05.01}%
  \BibitemOpen
  \bibfield  {author} {\bibinfo {author} {\bibfnamefont {O.~C.}\ \bibnamefont
  {Zienkiewicz}}, \bibinfo {author} {\bibfnamefont {R.~L.}\ \bibnamefont
  {Taylor}}, \ and\ \bibinfo {author} {\bibfnamefont {J.~Z.}\ \bibnamefont
  {Zhu}},\ }\href@noop {} {\emph {\bibinfo {title} {The Finite Element Method:
  Its Basis and Fundamentals}}},\ \bibinfo {edition} {6th}\ ed.\ (\bibinfo
  {publisher} {Butterworth-Heinemanm},\ \bibinfo {address} {Oxford},\ \bibinfo
  {year} {2005})\BibitemShut {NoStop}%
\bibitem [{\citenamefont {Sorensen}(1992)}]{soren92.01}%
  \BibitemOpen
  \bibfield  {author} {\bibinfo {author} {\bibfnamefont {D.~C.}\ \bibnamefont
  {Sorensen}},\ }\href@noop {} {\bibfield  {journal} {\bibinfo  {journal}
  {{SIAM} J. Matrix Anal. Appl.},\ }\textbf {\bibinfo {volume} {13}},\ \bibinfo
  {pages} {357} (\bibinfo {year} {1992})}\BibitemShut {NoStop}%
\bibitem [{\citenamefont {Kuzyk}\ and\ \citenamefont
  {Watkins}(2006)}]{kuzyk06.02}%
  \BibitemOpen
  \bibfield  {author} {\bibinfo {author} {\bibfnamefont {M.~G.}\ \bibnamefont
  {Kuzyk}}\ and\ \bibinfo {author} {\bibfnamefont {D.~S.}\ \bibnamefont
  {Watkins}},\ }\href@noop {} {\bibfield  {journal} {\bibinfo  {journal} {J.
  Chem Phys.},\ }\textbf {\bibinfo {volume} {124}},\ \bibinfo {pages} {244104}
  (\bibinfo {year} {2006})}\BibitemShut {NoStop}%
\bibitem [{\citenamefont {Szafruga}\ \emph {et~al.}(2010)\citenamefont
  {Szafruga}, \citenamefont {Kuzyk},\ and\ \citenamefont
  {Watkins}}]{szafr10.01}%
  \BibitemOpen
  \bibfield  {author} {\bibinfo {author} {\bibfnamefont {U.~B.}\ \bibnamefont
  {Szafruga}}, \bibinfo {author} {\bibfnamefont {M.~G.}\ \bibnamefont {Kuzyk}},
  \ and\ \bibinfo {author} {\bibfnamefont {D.~S.}\ \bibnamefont {Watkins}},\
  }\href@noop {} {\bibfield  {journal} {\bibinfo  {journal} {J. Nonl. Opt.
  Phys. \& Mat.},\ }\textbf {\bibinfo {volume} {19}},\ \bibinfo {pages} {379}
  (\bibinfo {year} {2010})}\BibitemShut {NoStop}%
\bibitem [{\citenamefont {Fu}\ \emph {et~al.}(2007){\natexlab{b}}\citenamefont
  {Fu}, \citenamefont {Lazaro}, \citenamefont {Hagan}, \citenamefont
  {Van~Stryland}, \citenamefont {Przhonska}, \citenamefont {Bondar},
  \citenamefont {Slominsky},\ and\ \citenamefont {Kachkovski}}]{fu07.02}%
  \BibitemOpen
  \bibfield  {author} {\bibinfo {author} {\bibfnamefont {J.}~\bibnamefont
  {Fu}}, \bibinfo {author} {\bibfnamefont {A.~P.}\ \bibnamefont {Lazaro}},
  \bibinfo {author} {\bibfnamefont {D.~J.}\ \bibnamefont {Hagan}}, \bibinfo
  {author} {\bibfnamefont {E.~W.}\ \bibnamefont {Van~Stryland}}, \bibinfo
  {author} {\bibfnamefont {O.~V.}\ \bibnamefont {Przhonska}}, \bibinfo {author}
  {\bibfnamefont {M.~V.}\ \bibnamefont {Bondar}}, \bibinfo {author}
  {\bibfnamefont {Y.~L.}\ \bibnamefont {Slominsky}}, \ and\ \bibinfo {author}
  {\bibfnamefont {A.~D.}\ \bibnamefont {Kachkovski}},\ }\href@noop {}
  {\bibfield  {journal} {\bibinfo  {journal} {J. Opt. Soc. Am. B},\ }\textbf
  {\bibinfo {volume} {24}},\ \bibinfo {pages} {67} (\bibinfo {year}
  {2007}{\natexlab{b}})}\BibitemShut {NoStop}%
\bibitem [{\citenamefont {Meyers}\ \emph {et~al.}(1994)\citenamefont {Meyers},
  \citenamefont {Marder}, \citenamefont {Pierce},\ and\ \citenamefont
  {Bredas}}]{meyer94.01}%
  \BibitemOpen
  \bibfield  {author} {\bibinfo {author} {\bibfnamefont {F.}~\bibnamefont
  {Meyers}}, \bibinfo {author} {\bibfnamefont {S.~R.}\ \bibnamefont {Marder}},
  \bibinfo {author} {\bibfnamefont {B.~M.}\ \bibnamefont {Pierce}}, \ and\
  \bibinfo {author} {\bibfnamefont {J.~L.}\ \bibnamefont {Bredas}},\
  }\href@noop {} {\bibfield  {journal} {\bibinfo  {journal} {J. Amer. Chem.
  Soc.},\ }\textbf {\bibinfo {volume} {116}},\ \bibinfo {pages} {10703}
  (\bibinfo {year} {1994})}\BibitemShut {NoStop}%
\bibitem [{\citenamefont {Marder}\ \emph {et~al.}(1991)\citenamefont {Marder},
  \citenamefont {Beratan},\ and\ \citenamefont {Cheng}}]{marde91.01}%
  \BibitemOpen
  \bibfield  {author} {\bibinfo {author} {\bibfnamefont {S.~R.}\ \bibnamefont
  {Marder}}, \bibinfo {author} {\bibfnamefont {D.~N.}\ \bibnamefont {Beratan}},
  \ and\ \bibinfo {author} {\bibfnamefont {L.-T.}\ \bibnamefont {Cheng}},\
  }\href@noop {} {\bibfield  {journal} {\bibinfo  {journal} {Science},\
  }\textbf {\bibinfo {volume} {252}},\ \bibinfo {pages} {103} (\bibinfo {year}
  {1991})}\BibitemShut {NoStop}%
\bibitem [{\citenamefont {Marder}\ \emph {et~al.}(1993)\citenamefont {Marder},
  \citenamefont {Perry}, \citenamefont {Bourhill}, \citenamefont {Gorman},
  \citenamefont {Tiemann},\ and\ \citenamefont {Mansour}}]{marde93.01}%
  \BibitemOpen
  \bibfield  {author} {\bibinfo {author} {\bibfnamefont {S.}~\bibnamefont
  {Marder}}, \bibinfo {author} {\bibfnamefont {J.~W.}\ \bibnamefont {Perry}},
  \bibinfo {author} {\bibfnamefont {G.}~\bibnamefont {Bourhill}}, \bibinfo
  {author} {\bibfnamefont {C.~B.}\ \bibnamefont {Gorman}}, \bibinfo {author}
  {\bibfnamefont {B.~G.}\ \bibnamefont {Tiemann}}, \ and\ \bibinfo {author}
  {\bibfnamefont {K.}~\bibnamefont {Mansour}},\ }\href@noop {} {\bibfield
  {journal} {\bibinfo  {journal} {Science},\ }\textbf {\bibinfo {volume}
  {261}},\ \bibinfo {pages} {186} (\bibinfo {year} {1993})}\BibitemShut
  {NoStop}%
\bibitem [{\citenamefont {Duan}\ \emph {et~al.}(2001)\citenamefont {Duan},
  \citenamefont {Huang}, \citenamefont {Cui}, \citenamefont {Wang},\ and\
  \citenamefont {Lieber}}]{duan2001indium}%
  \BibitemOpen
  \bibfield  {author} {\bibinfo {author} {\bibfnamefont {X.}~\bibnamefont
  {Duan}}, \bibinfo {author} {\bibfnamefont {Y.}~\bibnamefont {Huang}},
  \bibinfo {author} {\bibfnamefont {Y.}~\bibnamefont {Cui}}, \bibinfo {author}
  {\bibfnamefont {J.}~\bibnamefont {Wang}}, \ and\ \bibinfo {author}
  {\bibfnamefont {C.}~\bibnamefont {Lieber}},\ }\href@noop {} {\bibfield
  {journal} {\bibinfo  {journal} {Nature},\ }\textbf {\bibinfo {volume}
  {409}},\ \bibinfo {pages} {66} (\bibinfo {year} {2001})}\BibitemShut
  {NoStop}%
\bibitem [{\citenamefont {Miller}\ \emph {et~al.}(1984)\citenamefont {Miller},
  \citenamefont {Chemla}, \citenamefont {Damen}, \citenamefont {Gossard},
  \citenamefont {Wiegmann}, \citenamefont {Wood},\ and\ \citenamefont
  {Burrus}}]{miller1984band}%
  \BibitemOpen
  \bibfield  {author} {\bibinfo {author} {\bibfnamefont {D.}~\bibnamefont
  {Miller}}, \bibinfo {author} {\bibfnamefont {D.}~\bibnamefont {Chemla}},
  \bibinfo {author} {\bibfnamefont {T.}~\bibnamefont {Damen}}, \bibinfo
  {author} {\bibfnamefont {A.}~\bibnamefont {Gossard}}, \bibinfo {author}
  {\bibfnamefont {W.}~\bibnamefont {Wiegmann}}, \bibinfo {author}
  {\bibfnamefont {T.}~\bibnamefont {Wood}}, \ and\ \bibinfo {author}
  {\bibfnamefont {C.}~\bibnamefont {Burrus}},\ }\href@noop {} {\bibfield
  {journal} {\bibinfo  {journal} {Physical Review Letters},\ }\textbf {\bibinfo
  {volume} {53}},\ \bibinfo {pages} {2173} (\bibinfo {year}
  {1984})}\BibitemShut {NoStop}%
\bibitem [{\citenamefont {Hall}\ \emph {et~al.}(2010)\citenamefont {Hall},
  \citenamefont {Dao}, \citenamefont {Koike}, \citenamefont {Sasa},
  \citenamefont {Tan}, \citenamefont {Inoue}, \citenamefont {Yano},
  \citenamefont {Jagadish},\ and\ \citenamefont {Davis}}]{hall2010using}%
  \BibitemOpen
  \bibfield  {author} {\bibinfo {author} {\bibfnamefont {C.}~\bibnamefont
  {Hall}}, \bibinfo {author} {\bibfnamefont {L.}~\bibnamefont {Dao}}, \bibinfo
  {author} {\bibfnamefont {K.}~\bibnamefont {Koike}}, \bibinfo {author}
  {\bibfnamefont {S.}~\bibnamefont {Sasa}}, \bibinfo {author} {\bibfnamefont
  {H.}~\bibnamefont {Tan}}, \bibinfo {author} {\bibfnamefont {M.}~\bibnamefont
  {Inoue}}, \bibinfo {author} {\bibfnamefont {M.}~\bibnamefont {Yano}},
  \bibinfo {author} {\bibfnamefont {C.}~\bibnamefont {Jagadish}}, \ and\
  \bibinfo {author} {\bibfnamefont {J.}~\bibnamefont {Davis}},\ }\href@noop {}
  {\bibfield  {journal} {\bibinfo  {journal} {Applied Physics Letters},\
  }\textbf {\bibinfo {volume} {96}},\ \bibinfo {pages} {193117} (\bibinfo
  {year} {2010})}\BibitemShut {NoStop}%
\bibitem [{\citenamefont {Wang}\ \emph {et~al.}(2006)\citenamefont {Wang},
  \citenamefont {Hu}, \citenamefont {Beratan},\ and\ \citenamefont
  {Yang}}]{wangm06.01}%
  \BibitemOpen
  \bibfield  {author} {\bibinfo {author} {\bibfnamefont {M.}~\bibnamefont
  {Wang}}, \bibinfo {author} {\bibfnamefont {X.}~\bibnamefont {Hu}}, \bibinfo
  {author} {\bibfnamefont {D.~N.}\ \bibnamefont {Beratan}}, \ and\ \bibinfo
  {author} {\bibfnamefont {W.}~\bibnamefont {Yang}},\ }\href@noop {} {\bibfield
   {journal} {\bibinfo  {journal} {J. Am. Chem. Soc.},\ }\textbf {\bibinfo
  {volume} {128}},\ \bibinfo {pages} {3228} (\bibinfo {year}
  {2006})}\BibitemShut {NoStop}%
\bibitem [{\citenamefont {Shafei}\ \emph {et~al.}(2010)\citenamefont {Shafei},
  \citenamefont {Kuzyk},\ and\ \citenamefont {Kuzyk}}]{shafei10.01}%
  \BibitemOpen
  \bibfield  {author} {\bibinfo {author} {\bibfnamefont {S.}~\bibnamefont
  {Shafei}}, \bibinfo {author} {\bibfnamefont {M.~C.}\ \bibnamefont {Kuzyk}}, \
  and\ \bibinfo {author} {\bibfnamefont {M.~G.}\ \bibnamefont {Kuzyk}},\
  }\href@noop {} {\bibfield  {journal} {\bibinfo  {journal} {J. Opt. Soc Am.
  B},\ }\textbf {\bibinfo {volume} {27}},\ \bibinfo {pages} {1849} (\bibinfo
  {year} {2010})}\BibitemShut {NoStop}%
\bibitem [{\citenamefont {Kuzyk}(2006){\natexlab{b}}}]{kuzyk06.03}%
  \BibitemOpen
  \bibfield  {author} {\bibinfo {author} {\bibfnamefont {M.~G.}\ \bibnamefont
  {Kuzyk}},\ }\href@noop {} {\bibfield  {journal} {\bibinfo  {journal} {J. Chem
  Phys.},\ }\textbf {\bibinfo {volume} {125}},\ \bibinfo {pages} {154108}
  (\bibinfo {year} {2006}{\natexlab{b}})}\BibitemShut {NoStop}%
\end{thebibliography}

%

\end{document}